\documentclass[aps,amsmath,amssymb,fleqn]{revtex4}

\newcommand{\bed}{\[}
\newcommand{\eed}{\]}
\newcommand{\beq}{\begin{equation}}
\newcommand{\eeq}{\end{equation}}
\newcommand{\beqa}{\begin{eqnarray}}
\newcommand{\eeqa}{\end{eqnarray}}
\newcommand{\ket} [1] {\vert #1 \rangle}
\newcommand{\bra} [1] {\langle #1 \vert}
\newcommand{\braket}[2]{\langle #1 | #2 \rangle}

\newcommand{\mean}[1]{\langle #1 \rangle}

\newcommand{\gras}[1]{\bold{#1}}

\newcommand{\tr}{\mathop{\mathrm{tr}}}

\usepackage{graphicx}
\usepackage[mathscr]{eucal}
\usepackage{amsmath}
\usepackage{amssymb}
\usepackage{epstopdf}
\usepackage{epsfig}
\begin{document}

\title{Matrix Product States Algorithms and Continuous Systems}
\author{S. Iblisdir$^1$,  R. Or\'us$^{2}$ and J. I. Latorre$^1$}

\affiliation{
${}^1$ Dept. d'Estructura i Constituents de la Mat\`eria,
Universitat de Barcelona, 647 Diagonal, 08028 Barcelona, Spain \\
${}^2$ School of Physical Sciences, University of Queensland,
QLD 4072, Australia}

\date{\today}

\begin{abstract}
A generic method to investigate many-body continuous-variable systems is
pedagogically presented. It is based on the notion of matrix product states (so-called MPS) 
and the algorithms thereof. The method is quite versatile and can be applied to a wide variety of situations. As a first test, we show how it provides reliable results in the computation of fundamental properties of a chain of quantum harmonic oscillators achieving off-critical and critical  relative errors of the order of $10^{-8}$ and $10^{-4}$ respectively. Next, we use it to study the ground state properties of the quantum rotor model in one spatial dimension, a model that can be mapped to the Mott insulator limit of the 1-dimensional Bose-Hubbard model. At the quantum critical point, the central charge associated to the underlying conformal field theory can be computed with good accuracy by measuring the finite-size corrections of the ground state energy. Examples of MPS-computations both in the finite-size regime and in the thermodynamic limit are given. The precision of our results are found to be comparable to those previously encountered in the MPS studies of, for instance, quantum spin chains. Finally, we present a spin-off application: an iterative technique to efficiently get numerical solutions of partial differential equations of many variables. We illustrate this technique by solving Poisson-like equations with precisions of the order of $10^{-7}$. 
\end{abstract}

\maketitle

\section{Introduction and main results}

Generally, the behaviour of quantum many-body systems cannot be grasped with a purely analytical approach, i.e. merely from the properties of its constituents. As a result, studying them often calls for approximation schemes that exhibit two features: (i) to provide an effective description of the system based on the identification of relevant degrees of freedom, (ii) to prescribe a (numerical) method to predict mean values of observables from this description. One such scheme is that based on the Density Matrix Renormalisation Group (DMRG) \cite{whit92a,whit92b}, whose applications are countless (see for example \cite{scho04} and references therein for a review).

Recently, a lot of attention has been devoted to regarding DMRG from a 
quantum information perspective. The aim of this effort is twofold. First,  
to understand when and why DMRG succeeds in describing a many-body system. Second, to improve and generalise the method. This programme has met some success. We now understand better the relation between the difficulty to describe the state of a many-body system and the entanglement content of this state \cite{vida03,lato04,vers05}. It is also clearer why DMRG performs poorly in certain circumstances. The cases of spin chains with periodic boundary conditions \cite{vers04} or two-dimensional spin lattices \cite{vers04b}  are quite explicit in this respect. Many other interesting proposals have been made along those lines such as  methods to simulate the evolution (in real or imaginary time) of a many-body system, both in the finite-size regime and in the thermodynamic limit \cite{vida04,vida04b, vida04c, vida05, vida06, vers04c,banu05}, or methods to efficiently compute a partition function \cite{murg05}. The common notion underlying all these progress is that of 
matrix product states (MPS) \cite{mps1, mps2, mps3, mps4, mps5}, that is an ansatz that allows to efficiently describe the state of the system when it is finitely correlated.

The present work is devoted to the study of many-body system with continuous degrees of freedom (or continuous-variable quantum many-body systems), using the MPS description and the related algorithms. The main motivation of our work is to provide a tool that will eventually allow to study (1+1) dimensional quantum field theories (similar work in this direction has been already considered in \cite{hc5,hc4,hc6}). The structure of this paper is as follows.

\begin{itemize}

\item In section \ref{sec:trunc}, we describe the procedure of local truncation of the Hilbert spaces of each subsystem for a generic model that allows to use the MPS machinery.

\item In section \ref{sec:genmps}, we review several numerical techniques based on MPS in a detailed and pedagogical way, both in the finite regime and in the thermodynamic limit. 

\item In section \ref{sec:harmonic}, we apply our methods to study a chain of coupled harmonic oscillators. This model is a lattice version of a (1+1)-dimensional Klein-Gordon field. We compute the energy, the von Neumann entropies of blocks and correlators of the ground state. These computations allow us to proof our method.

\item Next, in section \ref{sec:quantumrotor}, we perform a detailed study of the 1-dimensional quantum rotor model. This model is known to be interesting in studying the Superfluid-Mott insulator phase transitions, or arrays of Josephson junctions.

\item In section \ref{sec:pdemps}, we point out that the MPS algorithms may have applications well beyond the problems for which they have been initially designed. Namely, we show how they could be used to efficiently compute numerical solutions for a wide variety of partial differential equations.

\item Finally, in section \ref{sec:conclusions} we summarize the conclusions of this work. 

\end{itemize}

We wish to notice that our proposal for treating continuous-variable systems with MPS is qualitatively different from the Gaussian MPS presented in \cite{Wolf}.

\section{Truncation of local Hilbert spaces in a generic model}\label{sec:trunc}

Let us consider an isolated system of $N$ identical particles living on some domain $\mathcal{D}$ and are characterised, each, by a continuous degree of freedom $x \in \mathcal{D}$. 

We suppose that the behaviour of this system is exactly described by the Schr\"odinger equation, with a time independent Hamiltonian $H_{\textrm{exact}}$. The ground state of this Hamiltonian reads
\beq
\ket{\Psi_{0,\textrm{exact}}}= \int dx_1 \ldots dx_N \Psi_{0,\textrm{exact}}(x_1, \ldots, x_N) \ket{x_1, \ldots, x_N}.
\eeq
The problem that we want to analyse is to find an approximation, 
$\ket{\Psi_0}$, of $\ket{\Psi_{0,\textrm{exact}}}$ in the sense that $E_0=\bra{\Psi_0} H_{\textrm{exact}} \ket{\Psi_0}$ is close to $E_{0,\textrm{exact}}=\bra{\Psi_{0,\textrm{exact}}} H_{{\rm exact}} \ket{\Psi_{0,\textrm{exact}}}$.  For this, we will make two approximations. First, let us expand $\Psi_0(x_1,\ldots,x_N)$ in orthogonal sets of functions associated with each variable: 
\beq\label{eq:exactexpansion}
\Psi_{0,\textrm{exact}}(x_1, \ldots , x_N)=\sum_{s_1 \ldots s_N=1}^{\infty} c(s_1 ,\ldots , s_N) \phi_{s_1}^{(1)}(x_1) \cdots \phi_{s_N}^{(N)}(x_N).
\eeq

We wish to consider two approximations of Eq.(\ref{eq:exactexpansion}). The first one consists in an appropriate truncation on the dimension of the local Hilbert spaces. To be precise, for each particle $k$, we truncate the associated basis to a finite set of $d$ orthonormal functions:
\beq\label{eq:approx1}
\Psi_{0,\textrm{exact}}(x_1 ,\ldots ,x_N)=\Psi_{0,d}(x_1, \ldots , x_N)+\textrm{Rest}_d.
\eeq
In this paper paper we shall always consider this local dimension $d$ to be the same for all particles.  Notice that it is possible to make a clever choice of the local $d$-dimensional subspaces according to the particularities of the specific situation. In fact, once the relevant truncated subspaces are chosen, it is possible to make them depend on variational parameters which can be optimally tuned in order to minimize the error of the approximations. Furthermore, as long as we restrict our computations to the truncated basis,  $H_{\textrm{exact}}$ can be replaced by the truncated Hamiltonian
\beq
H=\sum_{i_1, \ldots, i_N=1}^{d} \sum_{j_1, \ldots, j_N=1}^{d} 
\bra{\phi^{(1)}_{i_1} \otimes \cdots \otimes \phi^{(N)}_{i_N} } H_{\textrm{exact}}
\ket{\phi^{(1)}_{j_1} \otimes \cdots \otimes \phi^{(N)}_{j_N} }
\ket{\phi^{(1)}_{i_1} \otimes \cdots \otimes  \phi^{(N)}_{i_N} }
\bra{\phi^{(1)}_{j_1} \otimes \cdots \otimes \phi^{(N)}_{j_N} }.
\eeq 

Considering Hamiltonians of the form
\begin{equation}
\label{eq:genham}
H_{\textrm{exact}}=\frac{1}{2} \sum_{i=1}^N T_i+\sum_{i,j=1}^{N} V(x_i,x_j),
\end{equation}
we observe that $H$ can always be given the standard form 
\beq\label{eq:stdH}
H=\sum_{\gamma} \bigotimes_{i=1}^N h_i^{\gamma},
\eeq 
where $h_i^{\gamma}$ is a hermitean $d \times d$ matrix acting on the local Hilbert space at site $i$. 
This fact might not be obvious for some interactions (think of the Coulomb interaction for instance). We prove it as follows: suppose we have a potential term between two particles $(i,j)$ whose matrix elements read 
$V_{(i_1 i_2),(j_1 j_2)}=\int dx_1 dx_2 \phi^*_{i_1}(x_1) \phi^*_{j_1}(x_2)  V(x_1,x_2) 
\phi_{i_2}(x_1)  \phi_{j_2}(x_2)$. Performing the appropriate singular value decomposition, one gets
\bed
V_{(i_1 i_2),(j_1 j_2)}=\sum_{l=1}^{d^2} U_{(i_1 i_2),l} \; \Sigma_l \; W^*_{l,(j_1 j_2)},
\eed
which is of the desired form.

The second approximation is to assume that the coefficients $c(s_1,\ldots,s_N)$ are given by a product of finite-size matrices. At this point, it is possible to choose among several prescriptions which indeed involve different numerical optimization algorithms. We will review three of them.

\section{Matrix Product States and optimisation techniques}\label{sec:genmps}

Let us review briefly in this Section some of the main numerical optimization techniques based on MPS. Here we shall make a distinction among  three different cases, namely, those of periodic boundary conditions, open boundary conditions, and the thermodynamic limit. 

\subsection{Periodic boundary conditions}

Let us associate $d$ square matrices $A(k,s_k)$ for each particle $k=1 \ldots N$. These matrices are assumed to have dimensions $\chi \times \chi$. The periodic boundary condition ansatz consists in assuming that the coefficient $c(s_1, \ldots, s_N)$ can be written, in terms of them, as 
\beq\label{eq:approx2}
c(s_1, \ldots, s_N) = \tr(A(1,s_1) \cdots A(N,s_N)).
\eeq 

\begin{figure}[h]
\includegraphics[width=1\textwidth]{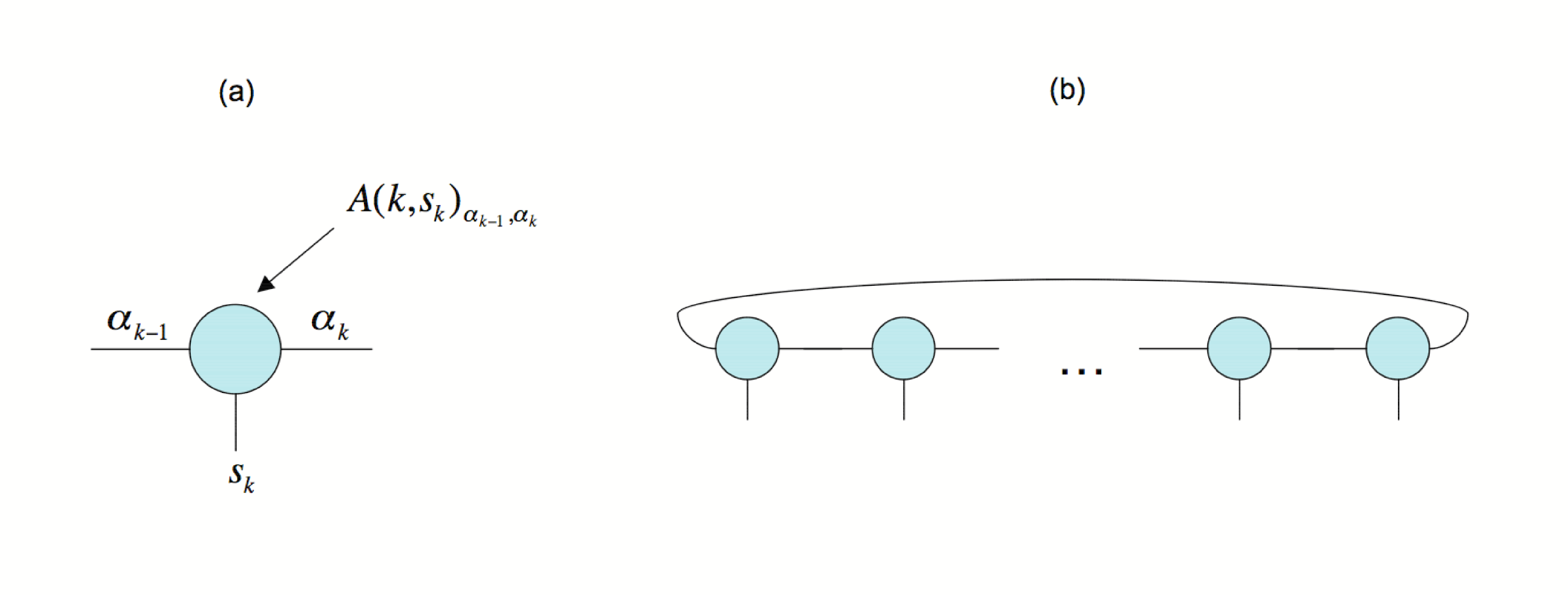}
\caption{(Color online) (a) Diagrammatic representation of the matrix element $A(k,s_k)_{\alpha_{k-1},\alpha_k}$ at site $k$. (b) Diagrammatic representation of the MPS structure of the coefficient $c(s_1, \ldots , s_N)$ (periodic boundary conditions).}
\label{matrixdiag}
\end{figure}

It is sometimes convenient to use diagrammatic notations for tensors, where open indices are represented by open legs, and contraction of indices is represented by gluing legs. The matrix elements $A(k,s_k)_{\alpha_{k-1},\alpha_k}$ at site $k$ and  the coefficient $c(s_1, \ldots , s_N)$ can then be represented as indicated in Fig.\ref{matrixdiag}. 
 
The representation given in Eq.(\ref{eq:approx2}) is efficient in the sense that it involves a relatively small number of independent  parameters \cite{vers04}. Indeed, $O(d^N)$ independent parameters are necessary in order to describe such a general state as Eq.(\ref{eq:approx1}). In contrast, the MPS description  requires only $O(Nd\chi^2)$ parameters, which for fixed $\chi$ is \emph{linear} in $N$. The approximation  (\ref{eq:approx2}) can be made arbitrarily precise upon taking $\chi$ sufficiently large. Actually taking $\chi=d^{\lfloor N/4 \rfloor}$ allows to describe any state of $N$ particles with local Hilbert spaces of dimension $d$ \cite{vers04} within this prescription. But remarkably, for many Hamiltonians of physical interest, $\ket{\Psi_{0,\textrm{exact}}}$ can be approximated accurately with quite a modest value of $\chi$ \cite{vers05}. To be precise, $\chi$ can be related to the entanglement entropy $S$ of a subsystem in such a way that, for the case considered here, $\chi \ge d^{S/2}$. Hence, MPS provide a faithful representation of the ground state of slightly-correlated quantum mechanical systems, as is the case of non-critical 1-dimensional quantum spin chains \cite{lato04}. 

\begin{figure}[h]
\includegraphics[width=1\textwidth]{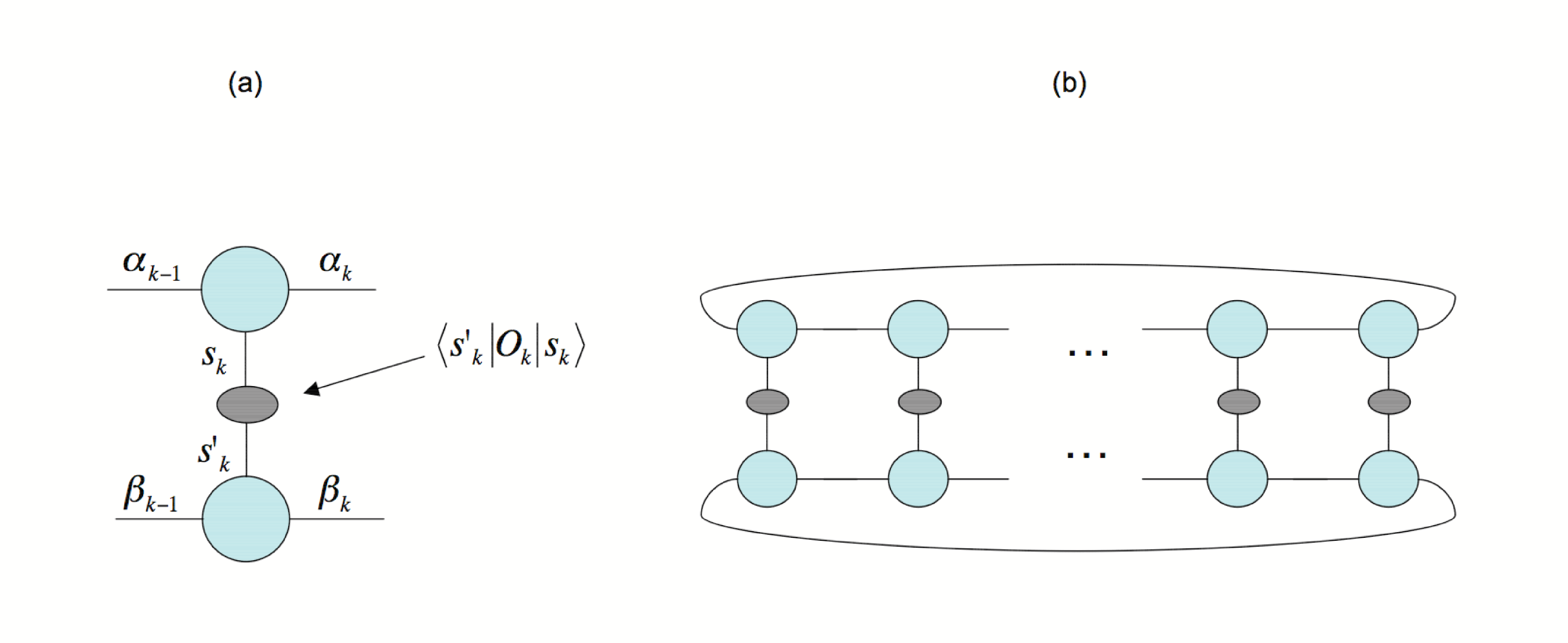}
\caption{(a) The diagrammatic representation of the transfer matrix $\widehat{O_k}$ at site $k$. (b) The diagrammatic representation of the mean value $\mean{O_1 \otimes \ldots \otimes O_N}=\tr(\widehat{O_1} \ldots \widehat{O_N})$, where the names of the indices are not explicitly indicated for simplicity.}
\label{pbcmv}
\end{figure}

In addition to providing an economical description of quantum states, MPS also allow for an efficient computation of expectation values of observables. Indeed suppose that we want to compute the mean value of a tensor product of local observables $\mean{O_1 \otimes \cdots \otimes O_N}$. One directly checks that the following identity holds
\beq
\mean{O_1 \otimes \cdots \otimes O_N}=\tr(\widehat{O_1} \cdots \widehat{O_N}),
\eeq
where $\widehat{O_k}=\sum_{s'_k,s_k=1}^{d} \bra{s'_k} O_k \ket{s_k} \bar{A}(k,s'_k) \otimes A(k,s_k)$ 
is a transfer matrix for the local operator $O_k$ at site $k$ ($\bar{A}$ denotes the conjugate matrix of $A$). These quantities are diagrammatically represented in Fig.\ref{pbcmv}, and it is easy to see that in this case the computation of the mean value of a tensor product of local operators requires of a computational time which grows as $O(N d \chi^5)$.
  
In order to compute an approximation to the ground state of the desired Hamiltonian operator $H$, the variational improvement now proceeds  as follows. First, one sets all matrices $A$ to some initial values. Then, one successively improves the matrices $A$ related to the particle $1$, then those of particle $2$, $\ldots$, then those of particle $N$, and then again. Sweeping sufficiently many times through the set $\{1, \ldots, N\}$, one converges to a stationary value of the approximated ground state energy $E_0$. The improvement of the matrices related to particle $k$ proceeds as follows. One computes the matrices $\mathcal{H}^{(k)}$ (the effective Hamiltonian matrix) and $\mathcal{N}^{(k)}$ (the normalisation matrix) such that 
\beqa
\mean{H} &=& \sum_{s'_k,s_k=1}^{d} \sum_{\beta_{k-1},\beta_k=1}^{\chi} \sum_{\alpha_{k-1},\alpha_k=1}^{\chi}
\bar{A}(k,s'_k)_{\beta_{k-1},\beta_k} 
\mathcal{H}^{(k)}_{(s'_k \beta_{k-1} \beta_k),(s_k \alpha_{k-1} \alpha_k)}
A(k,s_k)_{\alpha_{k-1},\alpha_k}, \\
\braket{\Psi_0}{\Psi_0} &=&  \sum_{s'_k,s_k=1}^{d} \sum_{\beta_{k-1},\beta_k=1}^{\chi} \sum_{\alpha_{k-1},\alpha_k=1}^{\chi}
\bar{A}(k,s'_k)_{\beta_{k-1},\beta_k} 
\mathcal{N}^{(k)}_{(s'_k \beta_{k-1} \beta_k),(s_k \alpha_{k-1} \alpha_k)}
A(k,s_k)_{\alpha_{k-1},\alpha_k}. 
\eeqa
The computation of both $\mathcal{H}^{(k)}$ and $\mathcal{N}^{(k)}$ follows the same rules as the computation of expected values in terms of transfer matrices, and is straightforward by using the tensor network diagrammatic representation from Fig.\ref{pbcmv} removing from the diagram the appropriate matrices at site $k$, given that $H$ has been brought to a normal form (\ref{eq:stdH}). Extremising $\mean{H}$ with the constraint $\braket{\Psi_0}{\Psi_0}=1$ with respect to the coefficients $\bar{A}(k,s_k)_{\alpha_{k-1},\alpha_k}$ leads to having to solve the (generalised) eigenvalue problem
\beq
 \sum_{s_k=1}^{d}  \sum_{\alpha_{k-1},\alpha_k=1}^{\chi}
 [
\mathcal{H}^{(k)}_{(s'_k \beta_{k-1} \beta_k),(s_k \alpha_{k-1} \alpha_k)}
- \lambda 
\mathcal{N}^{(k)}_{(s'_k \beta_{k-1} \beta_k),(s_k \alpha_{k-1} \alpha_k)}
]
A(k,s_k)_{\alpha_{k-1},\alpha_k}=0, \; \forall s'_k,\beta_{k-1},\beta_k,
\eeq
where a Lagrange multiplier $\lambda$ has been introduced.  The value of this Lagrange multiplier is actually the value of the ground state energy, as computed at the considered step of the algorithm.

%We shall see in what follows that this optimization procedure can be very much simplified (at the cost of a lost of precision depending on the model under study) by making use of open boundary conditions instead of periodic boundary conditions. Also, we shall see that other optimization methods may be applied when considering the thermodynamic limit, such as euclidean time evolution. 

\subsection{Open boundary conditions}

\begin{figure}[h]
\includegraphics[width=.5\textwidth]{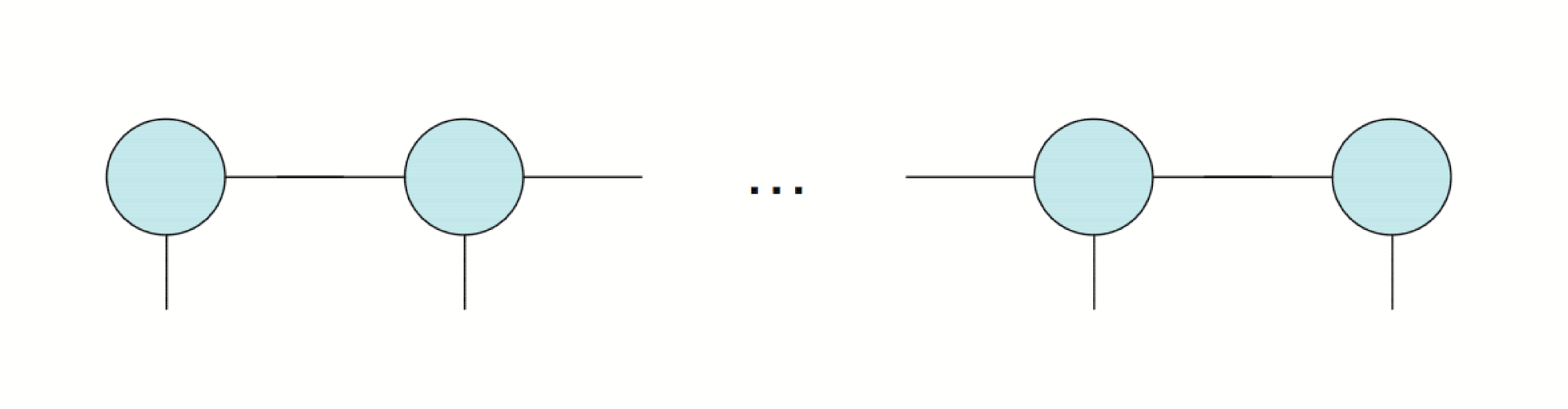}
\caption{(color online) Diagrammatic representation of the MPS structure of the coefficient $c(s_1, \ldots , s_N)$ in terms of open boundary conditions.}
\label{mpsobc}
\end{figure}

We assume again an ansatz of the form (\ref{eq:approx2}) but where the matrices now have different sizes, depending on the site to which they are related. The matrices $A(1,s)$ have size $1 \times \chi$, the matrices $A(k,s), k=2,\ldots,N-1$ have size $\chi \times \chi$, and the matrices $A(N,s)$ have size $\chi \times 1$. This matrix product structure can again be represented in terms of a diagrammatic tensor network, as shown in Fig.\ref{mpsobc}. 

As in the case of periodic boundary conditions, this MPS description is efficient as compared to the exact description of the $d^N$ coefficients $c(s_1, \ldots, s_N)$. Again, we need of $O(Nd\chi^2)$ parameters to describe the appropriate product of matrices, which can be made as precise as desired by increasing the parameter $\chi$. In this case, it is easy to see by considering the consecutive Schmidt decompositions of the system that if $\chi = d^{\lfloor N/2 \rfloor}$ then any state of $N$ particles of local dimensionality $d$ can be represented. 

Working with open boundary conditions is a stronger assumption than working with periodic ones. It is also a less physical approximation when studying systems described by a translationally invariant Hamiltonian. However, they allow to re-parametrise the MPS in a manner that considerably lowers the computational cost of each step of the optimisation procedure. This can be achieved thanks to the local gauge freedom existing in the definition (\ref{eq:approx2}). Let $T_k$ denote a $\chi \times \chi$ invertible matrix. The transformation $A(k,s) \to A(k,s)T_k , A(k+1,s) \to T_k^{-1} A(k+1,s)$ ($k < N$) leaves invariant the state described by the MPS.

We start with the site 1 and perform the singular value decomposition: 
\begin{equation}
A(1,s_1)_{1,\alpha_1}= \sum_{\lambda,\mu=1}^{{\rm min}(d,\chi)}
U^{(1)}_{s_1,\lambda} \Sigma^{(1)}_{\lambda,\mu} V^{(1)}_{\mu,\alpha_1}. 
\end{equation}
where by definition of singular value decompostion $\Sigma^{(1)}$ is a diagonal ${\rm min}(d,\chi) \times {\rm min}(d,\chi)$ matrix, and $U^{(1)}$ and $V^{(1)}$ satisfy $U^{(1)\dagger} U^{(1)}=\mathbb{I}_{{\rm min}(d,\chi)}$ and $V^{(1)} V^{(1)\dagger}=\mathbb{I}_{{\rm min}(d,\chi)}$. We then perform a gauge transformation 
\bed
A(1,s_1)_{1,\alpha_1} \to U(1,s_1)_{1, \lambda} \equiv U^{(1)}_{s_1,\lambda},
\eed
\bed
A(2,s_2)_{\alpha_1,\alpha_2} \to B(2,s_2)_{\lambda,\alpha_2}= \sum_{\mu=1}^{{\rm min}(d,\chi)} \sum_{\nu=1}^{\chi} \Sigma^{(1)}_{\lambda,\mu} V^{(1)}_{\mu,\nu} A(2,s_2)_{\nu,\alpha_2}. 
\eed
Observe that the dimensions of the matrices related to sites 1 and 2 may change along this transformation. Then we perform the further singular value decomposition for $B(2,s_2)$
\bed
B(2,s_2)_{\lambda,\alpha_2}=\sum_{\mu,\nu=1}^{{\rm min}(d^2,\chi)} U^{(2)}_{(s_2 \lambda),\mu} \Sigma^{(2)}_{\mu,\nu} V^{(2)}_{\nu,\alpha_2},
\eed
and make the gauge transformation
\bed
B(2,s_2)_{\lambda,\alpha_2} \to U(2,s_2)_{\lambda, \mu} \equiv U^{(2)}_{(s_2 \lambda),\mu},
\eed
\bed
A(3,s_3)_{\alpha_2,\alpha_3} \to B(3,s_3)_{\mu,\alpha_3}= \sum_{\nu=1}^{{\rm min}(d^2,\chi)} 
\sum_{\alpha_2=1}^{\chi}
\Sigma^{(2)}_{\mu,\nu} V^{(2)}_{\nu,\alpha_2} A(3,s_3)_{\alpha_2,\alpha_3}.
\eed

Iterating this procedure up to the site $k-1$, the state is represented through the product of matrices
\begin{equation}
U(1,s_1) \cdots U(k-1,s_{k-1}) B(k,s_k) A(k+1,s_{k+1}) \cdots A(N,s_N). 
\end{equation}
Then, one applies the same procedure, starting from site $N$, up to site $k+1$ this time, keeping at each step the matrix at the right of the singular decomposition. Eventually, our state is described by a product of matrices of the form
\begin{equation}
U(1,s_1) \cdots U(k-1,s_{k-1}) A(k,s_k) V(k+1,s_{k+1}) \cdots V(N,s_N).
\label{goodmps} 
\end{equation}

The principal virtue of this gauge is that $\braket{\Psi_0}{\Psi_0}$ now assumes a simple form, namely,
\begin{equation}
\braket{\Psi_0}{\Psi_0} =\sum_{s_k=1}^d \tr (A^*(k,s_k) A(k,s_k)),
\end{equation}
so that the corresponding normalization matrix $\mathcal{N}^{(k)}$ is now equal to the identity. This specific property of the chosen MPS representation for open boundary conditions allows a more straightforward numerical analysis of the corresponding variational optimization algorithm to compute good approximations of the ground state of Hamiltonians.  The variational improvement proceeds now by sweeping back and forth over the $N$ sets of matrices of the MPS optimizing over a particular site at each step. However, when optimizing over the set of matrices at site $k$, we shall consider now the MPS representation for open boundary conditions in terms of matrices $U(l,s_l)$ for sites $l=1,2,\ldots , k-1$, $V(l,s_l)$ for sites $l=k+1, k+2, \ldots , N$, and $A(k,s_k)$. The corresponding generalised eigenvalue problem is then transformed into a simple eigenvalue problem. Therefore, computing the ground state of the corresponding effective Hamiltonian $\mathcal{H}^{(k)}$ directly gives the variational ground state energy when optimizing over the site $k$, together with the corresponding well-normalized optimal matrix for that site. This can be very efficiently computed by means of large sparse-matrix techniques. Moving from site $k$ to site $k+1$ then proceeds by means of evaluating the new representation of the MPS in terms of $U(l,s_l)$ for sites $l=1,2,\ldots , k$, $V(l,s_l)$ for sites $l=k+2, k+2, \ldots , N$, and $A(k+1,s_{k+1})$, which only involves performing the appropriate singular value decomposition over the just-obtained optimal matrix at site $k$. As a matter of fact, the computation of $\mathcal{H}^{(k)}$ at each step (as well as the computation of expected values of local observables) is also very much simplified thanks to the considered structure of the open-boundary MPS, taking into account the normalization conditions from the left and from the right of the matrices forming the MPS.   

Finally, it is easy to see that considering the above procedures, together with efficient clever contractions of the different tensor networks that appear in the calculations, the computational running time of the variational optimization algorithm over MPS with open boundary conditions grows like $O(N^2 d\chi^3)$. 

Let us now present an alternative derivation of the MPS structure for open boundary conditions \cite{vida03}. If we perform the Schmidt decomposition between the local system $1$ and the remaining $N-1$, we can write the state as
\begin{equation}
|\Psi_{0} \rangle = \sum_{\alpha_1=1}^{{\rm min}(d,\chi)} \lambda(1)_{\alpha_1} |\tau^{(1)}_{\alpha_1}\rangle |\tau^{(2 \cdots n)}_{\alpha_1}\rangle \ , 
\label{smdec}
\end{equation}
where $\lambda(1)_{\alpha_1}$ are the Schmidt coefficients, $|\tau^{(1)}_{\alpha_1}\rangle$ and $|\tau^{(2 \cdots n)}_{\alpha_1}\rangle $ are the corresponding left and right Schmidt vectors. Expressing the left Schmidt vector in terms of the original local basis for system $1$, $\ket{\Psi_0}$ can then be written as
\begin{equation}
|\Psi_{0} \rangle = \sum_{s_1=1}^d \sum_{\alpha_1=1}^{{\rm min}(d,\chi)} \Gamma(1, s_1)_{1 \alpha_1} \lambda(1)_{\alpha_1} |\phi^{(1)}_{s_1} \rangle |\tau^{(2 \cdots n)}_{\alpha_1}\rangle,
\label{gamaequation}
\end{equation}
where $\Gamma(1, s_1)_{1\alpha_1}$ is the appropriate coefficients of the change of basis. That is, $ |\tau^{(1)}_{\alpha_1}\rangle = \sum_{s_1}  \Gamma(1, s_1)_{1 \alpha_1} |\phi^{(1)}_{s_1} \rangle $. At this point, we expand each Schmidt vector $|\tau^{(2 \cdots n)}_{\alpha_1}\rangle$ in the original local basis for system $2$, that is, 
\begin{equation}
 |\tau^{(2 \cdots n)}_{\alpha_1}\rangle = \sum_{s_2=1}^d |\phi^{(2)}_{s_2} \rangle |\omega^{(3 \cdots n)}_{\alpha_1 s_2} \rangle . 
 \label{cosita}
 \end{equation}
We now write the unnormalised quantum state  $|\omega^{(3 \cdots n)}_{\alpha_1 s_2} \rangle$  in terms of the at most $d^2$ eigenvectors of the joint reduced density matrix for systems $(3, \ldots, n)$, that is, in terms of the right Schmidt vectors $|\tau^{(3 \cdots n)}_{\alpha_2} \rangle $ of the particular bipartition between the first two local systems and the rest, together with the corresponding Schmidt coefficients $\lambda(2)_{\alpha_2}$: 
\begin{equation}
|\omega^{(3 \cdots n)}_{\alpha_1 s_2} \rangle = \sum_{\alpha_2=1}^{{\rm min}(d^2,\chi)} \Gamma(2, s_2)_{\alpha_1 \alpha_2} \lambda(2)_{\alpha_2} |\tau^{(3 \cdots n)}_{\alpha_2} \rangle  . 
\label{ecc}
\end{equation}
Replacing the last two expressions into Eq.(\ref{gamaequation}) we get 
\begin{equation}
|\Psi_{0} \rangle = \sum_{s_1, s_2 = 1}^d \sum_{\alpha_1=1}^{{\rm min}(d,\chi)} \sum_{\alpha_2=1}^{{\rm min}(d^2,\chi)} \Gamma(1, s_1)_{1 \alpha_1} \lambda(1)_{\alpha_1} \Gamma(2, s_2)_{\alpha_1 \alpha_2} \lambda(2)_{\alpha_2} |\phi^{(1)}_{s_1} \otimes \phi^{(2)}_{s_2} \rangle |\tau^{(3 \cdots n)}_{\alpha_2} \rangle \ . 
\label{iuid}
\end{equation}
Iterating the above procedure, we finally get a representation of the quantum state in terms of some tensors $\Gamma$ and some vectors $\lambda$: 
\begin{equation}
\label{gammalambda}
 |\Psi_{0,d} \rangle =
\sum_{\{s\}}\sum_{\{\alpha \}}\Gamma(1, s_1)_{1\alpha_1} \lambda(1)_{\alpha_1}
\Gamma(2, s_2)_{\alpha_1\alpha_2} \lambda(2)_{\alpha_2}
\dots \lambda(N-1)_{\alpha_{N-1}}\Gamma(N, s_N)_{\alpha_{N-1}1}\vert \phi^{(1)}_{s_1}  \phi^{(2)}_{s_2}  \cdots  \phi^{(N)}_{s_N} \rangle ,
 \end{equation} 
where the whole set of indices $\{s\}$ and $\{\alpha\}$ are added up to their respective allowed values. 
 
The above decomposition immediately provides the Schmidt vectors $\lambda$ of all the possible contiguous bipartitions of the system. In fact, Eq.(\ref{gammalambda}) is indeed a reparametrization of an open boundary condition MPS, in a gauge where
\begin{equation}
\label{agama}
A(k,s_k)_{\alpha_{k-1} \alpha_{k}} \equiv \Gamma(k, s_l)_{\alpha_{k-1} \alpha_k} \lambda(k)_{\alpha_k} .
\end{equation}

\begin{figure}[h]
\includegraphics[width=0.73\textwidth]{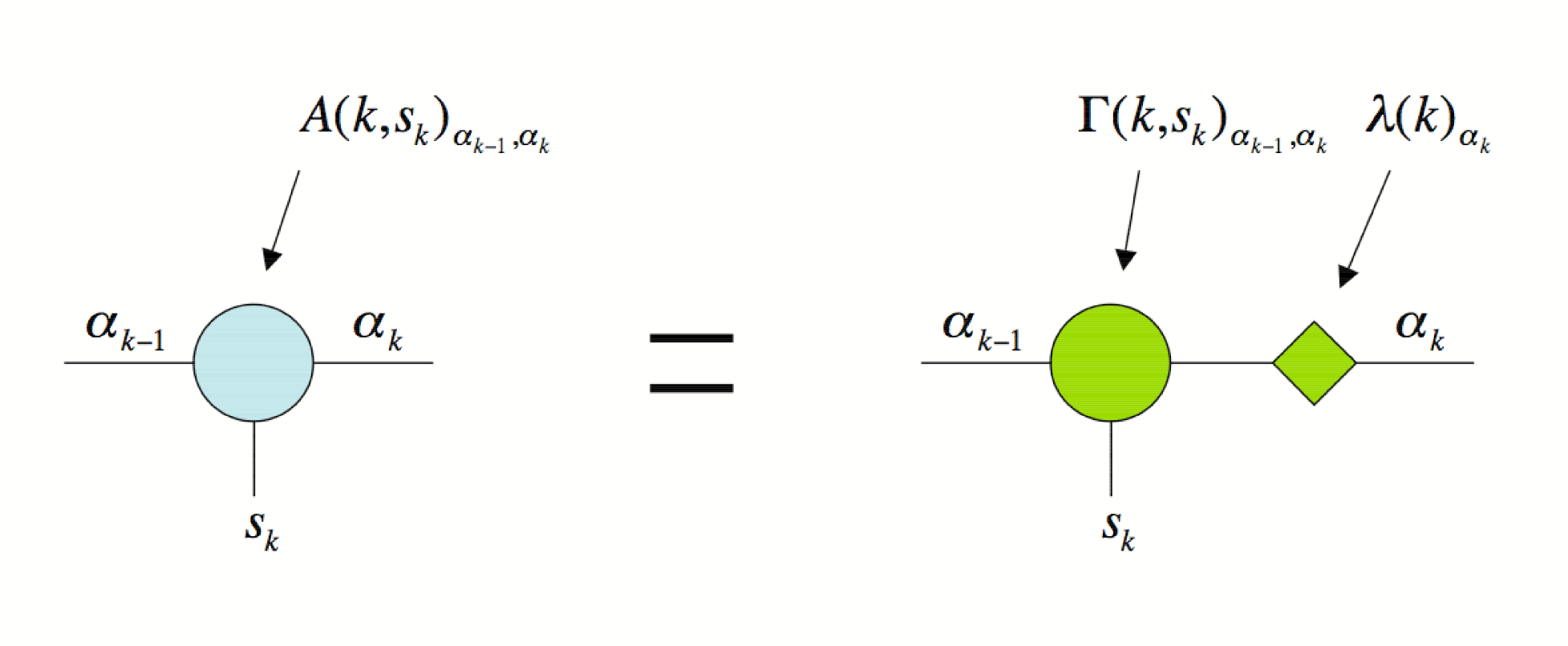}
\caption{(color online) The diagrammatic representation of the decomposition of $A(k,s_k)$ into a tensor $\Gamma(k,s_k)$ and a Schmidt coefficient $\lambda(k)$.}
\label{diaggam}
\end{figure}

This reparameterization of the MPS is diagrammatically represented in Fig.\ref{diaggam}. 
We also see that the maximum allowed rank of the different indices $\alpha_k$, $k = 1, \ldots , N-1$, is site-dependent, since the size of the Hilbert spaces considered when performing the consecutive  Schmidt decompositions depends on the site. 
In particular, we have that, at most,  $\alpha_k = 1, 2, \ldots, d^k$ for $k = 0, 1, \ldots, \lfloor N/2 \rfloor$, and $\alpha_k = 1, 2, \ldots, d^{(N-k)}$ for $k = N, N-1, \ldots, \lfloor N/2 \rfloor + 1$. The value of the true rank of the indices in the MPS comes dictated by the minimum between these quantities and the parameter $\chi$. 

In practice, when studying physically relevant states, many of the Schmidt coefficients for the different contiguous bipartitions of the system shall be (almost) equal to zero, depending on the particular state being considered. One is then interested in bounding the range of the indices by a number $\chi$, which now is clearly understood as a measure of the maximum contiguous bipartite entanglement in the system, since it is the maximum allowed Schmidt number for a contiguous bipartition in our MPS decomposition. Notice that this truncation in the Schmidt ranks imposes a strict restriction on the maximal amount of entanglement entropy that the ansatz can handle. To be precise, if $S = - {\rm tr}(\rho \log_2 \rho)$ is the von Neumann entropy corresponding to the reduced density matrix describing one of the subsystems of a bipartition of the quantum state into two contiguous pieces, then we see that $S \le \log_2 \chi$. This property makes these states to be {\it finitely-correlated}.

\subsection{The thermodynamic limit for translationally invariant systems}

In the case of a translationally invariant system, the techniques described in the foregoing analysis can be extended so as to consider an infinite number of particles, which allows to get rid of finite-size effects \cite{vida06}.  We now assume that all the properties of the state are defined in terms of a single set of matrices $A(s) \equiv \Gamma(s) \lambda$  which now do not depend on the position. 

It is indeed possible to perform the evolution of such an MPS in the thermodynamic limit both in real and euclidean time as driven by a translationally-invariant Hamitonian, as originally explained in \cite{vida06}. To this end, let us assume now that our Hamiltonian is a translationally invariant infinite sum of interaction terms which only involve nearest neighbour interactions, 
\begin{equation}
H = \sum_{k} h^{(k,k+1)} . 
\label{jami}
\end{equation}
The evolution operator $e^{-i H t}$ is now decomposed by means of a Suzuki-Trotter decomposition \cite{trotter} in terms of unitary gates acting only on two sites:
\begin{equation}
U^{(k,k+1)} = e^{-i h^{(k,k+1)} \delta t}, 
\label{deltat}
\end{equation}
where $\delta t \ll 1$, and which have to be applied $t/\delta t$ times. Arranging the above gates into two mutually non-conmuting pieces
\begin{eqnarray}
U(AB) &\equiv& \otimes_k U^{(2k,2k+1)} \nonumber \\
U(BA) &\equiv& \otimes_k U^{(2k-1,2k)}   , 
\label{unitarias}
\end{eqnarray}
we can compute the evolution in the MPS by slightly breaking the translational invariance as follows: since the set of gates in $U(AB)$ can be applied at once (the same holds for those in $U(BA)$), we consider an MPS invariant under shifts of two lattice sites, so that 
\begin{eqnarray}
\Gamma(2k,s_{2k}) &=& \Gamma(A,s_A) \nonumber \\
\Gamma(2k+1,s_{2k+1}) &=& \Gamma(B,s_B) 
\end{eqnarray}
and 
\begin{eqnarray}
\lambda(2k) &=& \lambda(A) \nonumber \\
\lambda(2k+1) &=& \lambda(B)  , 
\end{eqnarray}
for all the possible values of $k$. The updating of the MPS along the time evolution is then dictated by sequentially updating matrices for the $A$ and $B$ sites according to the application of the gates in Eq.(\ref{unitarias}). 
As explained in \cite{vida06}, the updating rule for the matrices of the MPS once a two-particle unitary gate is applied can be done by means of the appropriate singular value decomposition followed by a truncation up to $\chi$ of the rank of the common index that joins the matrices at the two sites. 

\begin{figure}[h]
\includegraphics[width=1\textwidth]{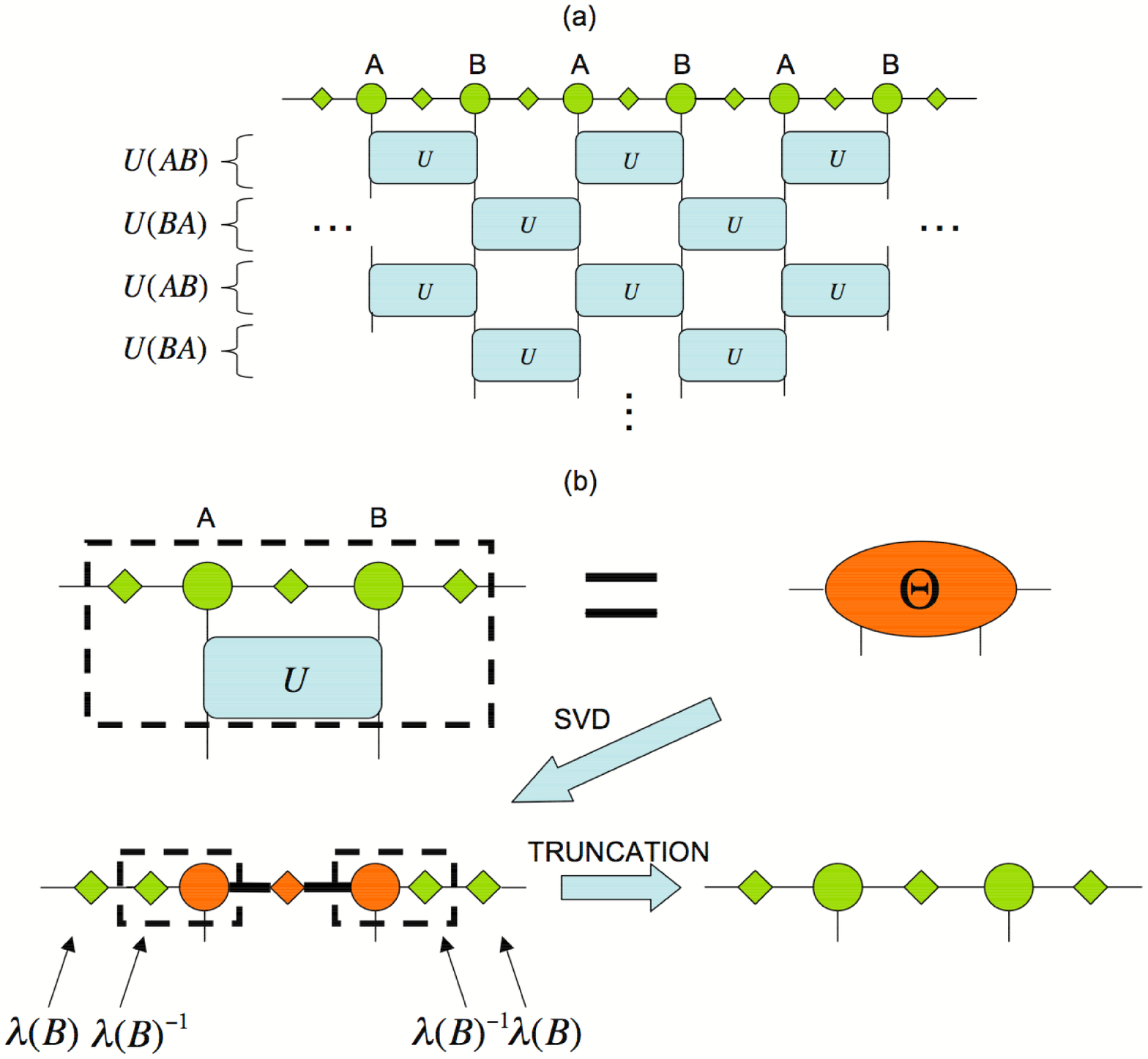}
\caption{(color online) (a) The sequential application of two-particle unitary gates to the MPS structure. (b) The updating rule for the MPS in the thermodynamic limit under the action of a two-particle gate follows the singular value decomposition (SVD) of the composed tensor, and a local truncation up to $\chi$ of the common index between the two particles. After the singular value decomposition, the vectors $\lambda$ are properly inserted by hand on the two sides in order to preserve the MPS structure, which also changes the values of the tensors at $A$ and $B$ by a factor of $\lambda^{-1}$.}
\label{evolution}
\end{figure}

Let us be more precise with the above statement. In the situation in which a two-body gate $U^{(A,B)}$ from (\ref{deltat}) is applied over two consecutive sites $A$ and $B$, the updating procedure starts with the following tensor contraction:
\begin{equation}
\Theta(s'_A, s'_B) \equiv \sum_{s_A = 1}^d \sum_{s_B = 1}^d U^{(A,B)}(s'_A, s'_B; s_A, s_B) \lambda(B) \Gamma(A, s_A) \lambda(A) \Gamma(B,s_B) \lambda(B). 
\end{equation}
By performing the singular value decomposition of the above tensor with respect to the left and right indices, we get
\begin{equation}
\Theta(s'_A, s'_B) = A(s'_A) \lambda'(A) B(s'_B) = \lambda(B) \Gamma'(A,s'_A) \lambda'(A) \Gamma'(B,s'_B) \lambda(B), 
\end{equation}
where $\Gamma'(A,s'_A) \equiv \lambda^{-1}(B) A(s'_A)$ and $\Gamma'(B, s'_B) \equiv B(s'_B) \lambda^{-1}(B)$ respectively correspond to the updated new matrices for sites $A$ and $B$. The last step in the updating procedure comes dictated by a truncation in the eigenvalues $\lambda'(A)$, in such a way that the final size of the updated matrices is again $\chi \times \chi$. 
The sequential application of gates and the truncation scheme are diagramatically represented in Fig.\ref{evolution}. 

Following this updating rule, it is possible to compute the evolution in time of any MPS in the thermodynamic limit as driven by a Hamiltonian such as the one from Eq.(\ref{jami}) in $O(d^3 \chi^3)$ time. Also, by performing evolution in euclidean time and taking very good care of the normalization of the wave function at each computational step, it is possible to find MPS approximations to the ground state of a wide variety of interacting Hamiltonians such as the ones considered in this paper. Note the difference between this optimization procedure, based on euclidean time evolution in the thermodynamic limit, and the ones described for finite systems in terms of local optimizations of the matrices by performing sweeps over the different particles, which involved the solution to (generalised) eigenvalue problems.

\section{Benchmark: the harmonic chain}\label{sec:harmonic}

As an illustration of the above methods, we consider an open chain of spin$-0$ particles interacting via the Hamiltonian
\beq\label{eq:hamiltonoh}
H_{\textrm{exact}}= \frac{1}{2} \sum_{i=1}^N p_{i}^{2}+ \mu \sum_{i=1}^{N} x_i^2+ 
\Lambda \sum_{i=1}^{N-1} (x_{i}-x_{i+1})^{2},
\eeq
corresponding to the well-known case of interacting harmonic systems \cite{jens,hc1,hc2,hc3,hc6,hc4}, and which is critical in the thermodynamic limit for $\mu = 0$. 
A remarkable feature of this Hamiltonian is obtained upon considering the continuous limit
\beq
\lim_{N \to \infty} \lim_{\Lambda \to \infty} H_{\textrm{exact}} \approx \int dq \; (\partial_{\nu}\partial^{\nu} + m^2 )\varphi(q),
\eeq
where we have made the identification $i \to q$, $x \to \varphi$, $\mu \to m^2$, and which corresponds to the usual Klein-Gordon Hamiltonian of a free spinless field $\varphi$. 

The ground state of $H_{\textrm{exact}}$ and its associated energy, $\ket{\Psi_{0,\textrm{exact}}}$ and $E_{0,\textrm{exact}}$, can be analytically computed \cite{sred93}. Rewriting $H_{\textrm{exact}}$ as
\beq
H_{\textrm{exact}}=\frac{1}{2} \sum_{i=1}^N p_i^2 + \sum_{i,j=1}^{N} x_i K_{i,j} x_j,
\eeq
we have $\Psi_{0,\textrm{exact}}(x_1, \ldots , x_N)= C e^{- \sum_{i,j=1}^N x_i (\sqrt{K})_{ij} x_j}$, $C$ being a normalization constant, and $E_{0,\textrm{exact}}=\frac{1}{2} \tr (\sqrt{K})$. This result will allow us to estimate the accuracy of our calculations.

Let us now apply the methods discussed in the previous sections and search for an MPS approximation to the ground state. As basis functions, we have chosen to use, for all particles,  the (truncated) spectrum of a single harmonic oscillator. That is,
\beq\label{eq:approxhoh}
\phi_{j}(x)=C_{j} \; e^{-a^2 x^2/2} \;  \mathcal{H}_{j}(ax), \hspace{1cm} j=0 \ldots d-1.
\eeq
In this expression, $C_j$ is a normalisation constant, $\mathcal{H}_j$ denotes the degree-$j$ Hermite polynomial, and $a=(2 \mu)^{1/4}$ (notice that we could also keep $a$ as an external variational parameter). With this choice $H_{\textrm{exact}}$ can be approximated by the truncated Hamiltonian
\beq
H=\frac{1}{2}\sum_{k=1}^{N} \tau^{(k)}+ (\Lambda+\mu) (\zeta^{(1)}+\zeta^{(N)})+
(2 \Lambda+\mu) \sum_{k=2}^{N-1} \zeta^{(k)}-2 \Lambda \sum_{k=1}^{N-1} \eta^{(k)} \eta^{(k+1)},
\eeq
where $\tau,\zeta$ and $\eta$ are hermitean matrices defined as
\beqa
\tau_{m,n} &=& - \int \; dx \; \phi_{m}(x) \frac{d^2}{dx^2} \phi_{n}(x) \nonumber \\
\zeta_{m,n} &=& \int \; dx \; \phi_{m}(x) \; x^2 \; \phi_{n}(x) \nonumber \\
\eta_{m,n}  &=&  \int \; dx \; \phi_{m}^{(i)}(x) \; x \; \phi_{n}(x),
\eeqa
where $m,n=0 \ldots d-1$. Analytical expressions for the entries of the matrices $\tau,\zeta$ and $\eta$ are straightforwardly derived from the elementary properties of Hermite polynomials (see \cite{abra72}). The Hamiltonian (\ref{eq:approxhoh}) is now of the form (\ref{eq:stdH}), so that its properties can be investigated with the algorithms of the previous sections. To be precise, we show here the performance of the finite-size numerical optimization algorithms for a system of $N=30$ sites and open boundary conditions, for several values of the external parameters. Notice that our simulations allow to change both $d$ and $\chi$ independently. 

\subsection{Relative error of the ground state energy}

We have performed computations of an approximated ground state of the system and compared the obtained results with the exact ones, for several values of $d$ and $\chi$. Our comparison is made precise by means of calculating the relative error of the ground state energy, as measured by the quantity
\begin{equation}
\delta \equiv \left| \frac{E_0 - E_{0,{\rm exact}}}{E_{0,{\rm exact}}} \right|. 
\end{equation}
In the off-crtical regime ($\mu \ne 0$), it is seen in table \ref{tab:hc} that the relative error of the ground state energy decreases exponentially with $d$ down to $10^{-8}$, while it remains very stable for the considered values of $\chi$. On the other hand, the exponential behaviour of the relative error as a function of $d$ becomes weaker when close to the criticality, as can be observed also in table \ref{tab:hc}, where it decreases up to $10^{-4}$. Let us remark that the precisions that we find in our calculations are in agreement and can compete with those found in previous numerical analysis of the harmonic chain \cite{hc1,hc2,hc3,hc4}.

\begin{table}
\begin{center}
\begin{tabular}{|c|c|c|c|c|}
\hline
 $N$ & $d$ & $\chi$ & $\delta (\mu = 1)$ & $\delta (\mu = 10^{-6})$\\
\hline
\hline
$30$ & $4$ & $5$&$ 8.27 \times 10^{-3} $ & $ 5.38 \times 10^{-3} $\\
$30$ & $4$ & $10$&$ 8.27 \times 10^{-3} $ & $ 5.23 \times 10^{-3} $\\
$30$ & $4$ & $15$&$ 8.27 \times 10^{-3}$ & $ 5.23 \times 10^{-3}$\\
$30$ & $4$ & $20$&$ 8.27 \times 10^{-3}$ & $ 5.23 \times 10^{-3}$\\
$30$ & $4$ & $25$&$ 8.27 \times 10^{-3}$ & $ 5.23 \times 10^{-3}$\\
\hline
$30$ & $6$ & $5$&$ 5.95 \times 10^{-4} $ & $ 3.07 \times 10^{-3}$\\
$30$ & $6$ & $10$&$ 5.95 \times 10^{-4}$ & $ 1.71 \times 10^{-3}$\\
$30$ & $6$ & $15$&$ 5.95 \times 10^{-4}$ & $ 1.68 \times 10^{-3}$\\
$30$ & $6$ & $20$&$ 5.95 \times 10^{-4}$ & $ 1.67 \times 10^{-3}$\\
$30$ & $6$ & $25$&$ 5.95 \times 10^{-4}$ & $ 1.67 \times 10^{-3}$\\
\hline
$30$ & $8$ & $5$&$ 4.03 \times 10^{-5} $ & $ 3.01 \times 10^{-3}$\\
$30$ & $8$ & $10$&$ 4.01 \times 10^{-5}$ & $ 1.04 \times 10^{-3}$\\
$30$ & $8$ & $15$&$  4.01 \times 10^{-5}$& $ 8.65 \times 10^{-4}$\\
$30$ & $8$ & $20$&$ 4.01 \times 10^{-5}$ & $ 8.48 \times 10^{-4}$\\
$30$ & $8$ & $25$&$ 4.01 \times 10^{-5}$ & $ 8.46 \times 10^{-4}$\\
\hline
$30$ & $10$ & $5$&$ 2.82 \times 10^{-6} $ & $ 2.94 \times 10^{-3}$\\
$30$ & $10$ & $10$&$ 2.63 \times 10^{-6}$ & $ 9.36 \times 10^{-4}$\\
$30$ & $10$ & $15$&$ 2.63 \times 10^{-6}$ & $ 6.46 \times 10^{-4}$\\
$30$ & $10$ & $20$&$ 2.63 \times 10^{-6}$ & $ 5.34 \times 10^{-4}$\\
$30$ & $10$ & $25$&$ 2.63 \times 10^{-6}$ & $ 5.24 \times 10^{-4}$\\
\hline
$30$ & $12$ & $5$&$ 3.63 \times 10^{-7} $ & $3.06 \times 10^{-3} $\\
$30$ & $12$ & $10$&$ 1.73 \times 10^{-7}$ & $ 9.47 \times 10^{-4}$\\
$30$ & $12$ & $15$&$ 1.72 \times 10^{-7}$ & $ 5.72 \times 10^{-4}$\\
$30$ & $12$ & $20$&$ 1.72 \times 10^{-7}$ & $ 4.4 \times 10^{-4}$\\
$30$ & $12$ & $25$&$ 1.72 \times 10^{-7}$ & $ 3.70 \times 10^{-4}$\\
\hline
$30$ & $14$ & $5$&$ 2.05 \times 10^{-7} $ & $ 2.98 \times 10^{-3}$\\
$30$ & $14$ & $10$&$ 1.42 \times 10^{-8}$ & $ 9.07 \times 10^{-4}$\\
$30$ & $14$ & $15$&$ 1.41 \times 10^{-8}$ & $ 5.60 \times 10^{-4}$\\
$30$ & $14$ & $20$&$ 1.41 \times 10^{-8}$ & $ 3.75 \times 10^{-4}$\\
$30$ & $14$ & $25$&$ 1.41 \times 10^{-8}$ & $ 3.21 \times 10^{-4}$\\
\hline
\end{tabular}
\end{center}
\caption{Performances of the MPS algorithm for computing the ground state energy of the harmonic chain, both in the off-critical regime ($\mu = 1$) and close to criticality ($\mu = 10^{-6}$). $N$ denotes the number of variables, $d$ denotes the number of local levels, $\chi$ is the dimension of the matrices of the MPS ansatz,  and $\delta$ the relative error.}\label{tab:hc}
\end{table}

\subsection{Off-critical correlator}

\begin{figure}[h]
\includegraphics[width=.51\textwidth,angle=-90]{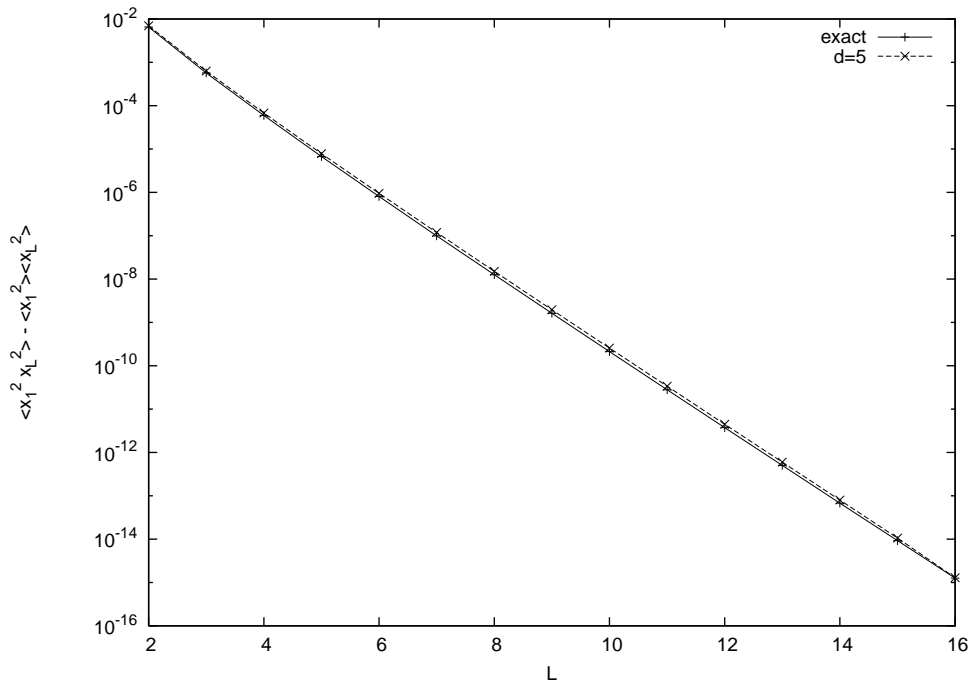}
\caption{Off-critical correlator $C(L) = \mean{ x^2_1 \ x^2_L} -  \mean{ x^2_1} \mean{x^2_L}$ for $N=30$ and $\mu =1$. The numerical computations have been done for $N=30$, $\chi = 30$, and $d=5$. Very good agreement with the exact value is observed to appear.}
\label{correl}
\end{figure}

Our numerical calculations allow us to compare the performance of the simulation technique by means of considering the correlation functions of the system. To be precise, we have numerically computed the two-point correlation function 
\begin{equation}
C(L) =\mean{x^2_1 \ x^2_L}- \mean{x^2_1} \mean{x^2_L}, 
\end{equation} 
for $N=30$, $\chi  = 30$ and $d=5$, and compared it to its exact value. It is a fact that, when close to criticality, the correlators are not well reproduced by means of our MPS techniques (which inherently make use of a finite correlation length). However, away from criticality ($\mu = 1$), our algorithms provide good approximations of the correlation functions, as can be seen in Fig. \ref{correl}, where it is shown that the numerical results obtained for $C(L)$ agree very well with the exact value. 

\subsection{Absolute error of the entanglement entropy}

\begin{figure}[h]
\includegraphics[width=.7\textwidth]{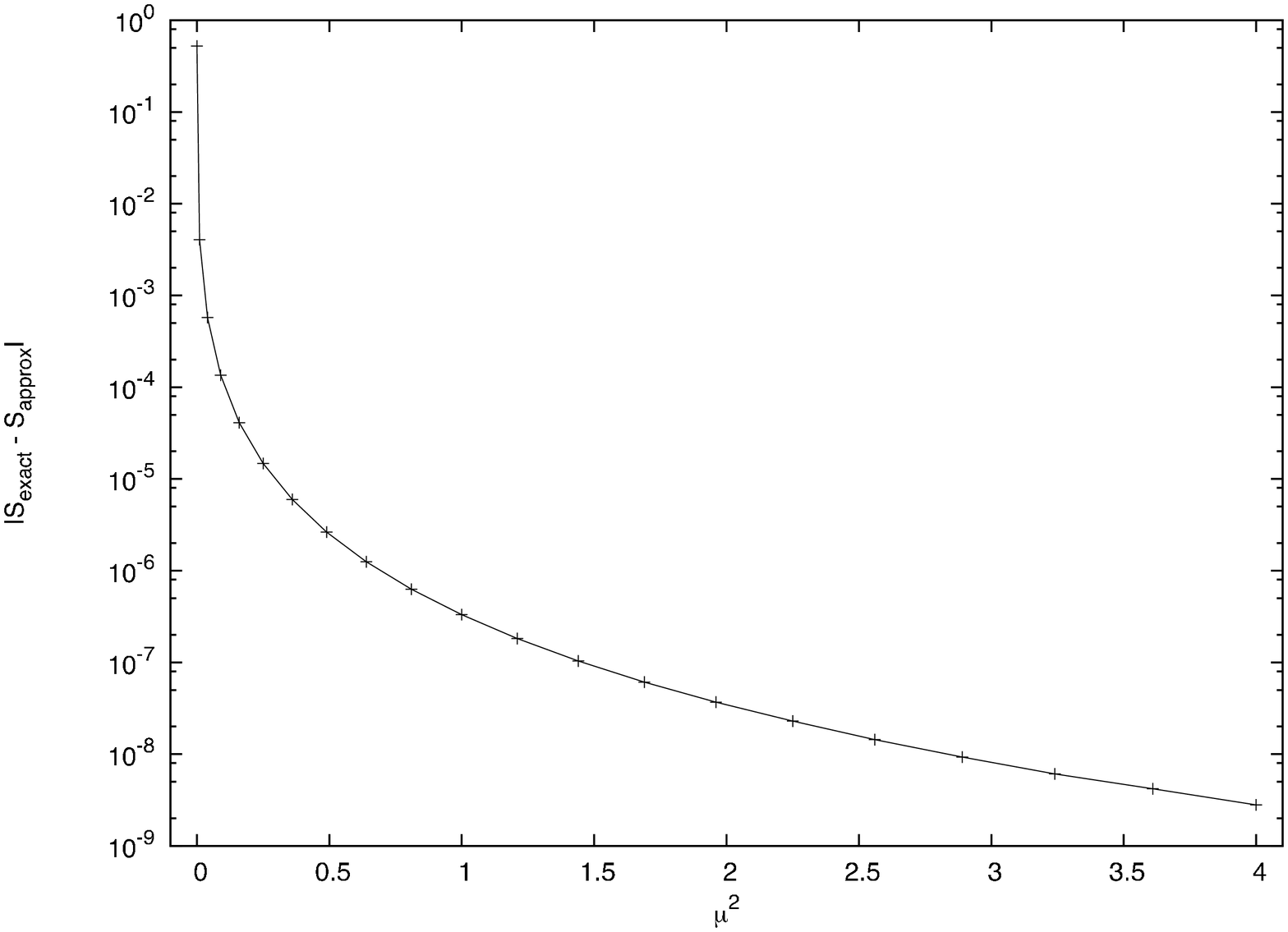}
\caption{Absolute error in the entanglement entropy for the harmonic chain for $N=30$, $d=14$, $\chi = 10$ and $\Lambda = 0.5$, as a function of $\mu^2$.}
\label{entroperr}
\end{figure}

\begin{figure}[h]
\includegraphics[width=.51\textwidth,angle=-90]{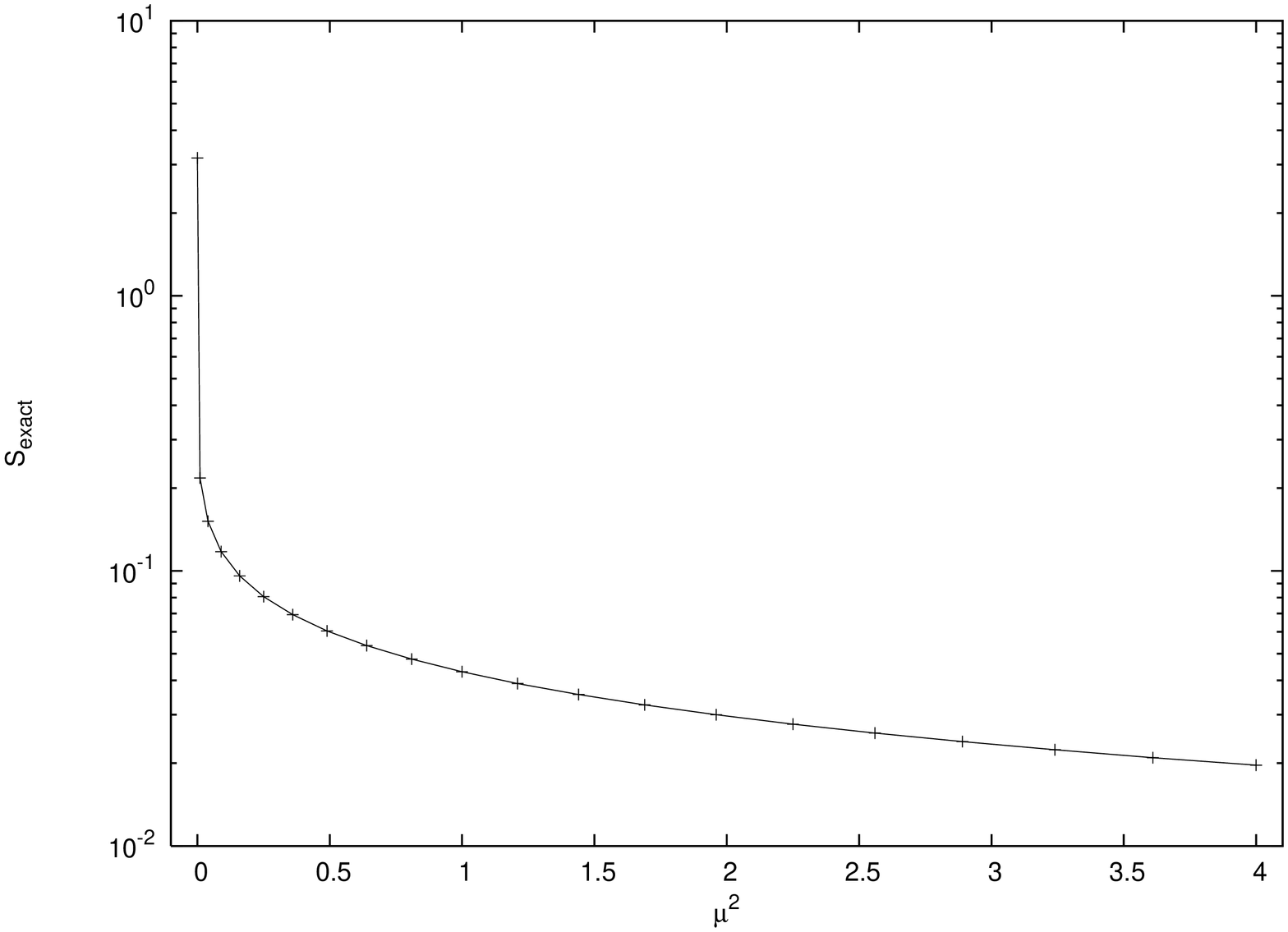}
\caption{Exact entanglement entropy for the harmonic chain for $N=30$ and $\Lambda = 0.5$, as a function of $\mu^2$.}
\label{exactentropy}
\end{figure}

Recently, a lot of attention has been devoted to computing the entanglement present in Gaussian systems (see for instance \cite{jens} and references therein). This motivated us to look at the entanglement entropy $S = -{\rm tr} (\rho \log_2 \rho)$ of the ground state reduced density matrix $\rho$ of half the open chain, as a further test of our method. We have computed $S$ for $\chi=10$, $d=14$ and $\Lambda = 0.5$, as a function of the parameter $\mu^2$, and compared it to the exact case. As can be seen in Fig.\ref{entroperr}, the absolute error decreases superexponentially when departing from the critical region, easily achieving values of the order of $10^{-9}$ when being sufficiently away from criticality, and giving small relative deviations from the exact entropy in Fig.\ref{exactentropy}. 

\section{The 1-dimensional quantum rotor}\label{sec:quantumrotor}

As a further example of the application of the techniques presented in the previous sections, we consider here the properties of the ground state of the so-called quantum rotor model in 1 spatial dimension. This continuous-variable model is defined by means of the Hamiltonian
\begin{equation}
H_{{\rm exact}} = \sum_{k=1}^N \left(-2 J \cos{(\theta_k - \theta_{k+1})} - \frac{U}{2} \frac{\partial^2}{\partial \theta_k^2}\right), 
\label{qrotor}
\end{equation}
where the angles $\theta_k \in [0,2\pi ) \ \forall k$, and $U, J$ are external parameters. The above Hamiltonian accurately describes the behaviour within the Mott-insulator regime of the 1-dimensional Bose-Hubbard model, where small fluctuations around the mean number of atoms per site can be considered \cite{juanjo,fisher}, hence providing an approximate description in a specific regime of the behavior of cold atoms in 1-dimensional optical lattices. This model has also met success in reproducing the different properties regarding the tunneling of Cooper pairs between different superconducting islands in arrays of Josephson junctions \cite{josephson}. From a technical point of view, note that the model is invariant under global rotations of all the angles, that is, under the global $U(1)$ symmetry group. A suitable tuning of the parameters defining the model describes then in the thermodynamic limit the evolution towards a quantum phase transition, whose $U(1)$-invariant critical point can be described in terms of the conformal field theory of a free scalar field with central charge $c=1$, as proven in \cite{affleck}. 

Let us perform now a suitable truncation on the local dimension of the different Hilbert spaces as follows.  We choose the individual particle basis of plane waves
\begin{equation}
\phi^{(k)}_{s_k} (\theta_k) = \frac{1}{\sqrt{2 \pi}} e^{i s_k \theta_k} , 
\end{equation}
with $s_k = -(d-1)/2, \ldots , 0 , \ldots , (d-1)/2$, for all the possible $k$. Restricted to this truncated basis, the Hamiltonian from Eq.(\ref{qrotor}) can be expressed as 
\begin{equation}
H = \sum_{k=1}^N \left( -2 J \left( V^{(k)} V^{(k+1)} + W^{(k)} W^{(k+1)} \right) + \frac{U}{2} T^{(k)} \right). 
\label{truncated}
\end{equation}
The matrix elements of operators $V$, $W$ and $T$ adopt the simple form
\begin{eqnarray}
V_{m,n} &=& \frac{1}{2} \left( \delta_{m-1,n}  + \delta_{m+1,n} \right) \nonumber \\
W_{m,n} &=& \frac{1}{2i} \left( \delta_{m-1,n} - \delta_{m+1,n} \right) \nonumber \\
T_{m,n} &=& n^2 \delta_{m,n} . 
\end{eqnarray}
Note that for $d=2$ we have that $V = \frac{1}{2}\sigma_x$, $W = \frac{1}{2} \sigma_y$ and $T = (\sigma_z)^2$. The truncated Hamiltonian given in Eq.(\ref{truncated}) for $d=2$ then represents a very specific case of the spin-$1/2$ $XY$ quantum spin chain, which was already known to be an approximation of the 1-dimensional Bose-Hubbard model in the Mott-insulator regime \cite{lato04}. 

We have performed a variety of calculations over the above truncated Hamiltonian for several values of the local dimension $d$ of the Hilbert space, by means of the MPS-like algorithms described at the beginning of this work. Our results extend those computed in \cite{juanjo}, and can be understood in terms of the underlying quantum phase transition that the model undergoes in the thermodynamic limit. 

\subsection{Ground state energy and entanglement entropy of half a system}

\begin{figure}
\includegraphics[width=.7\textwidth]{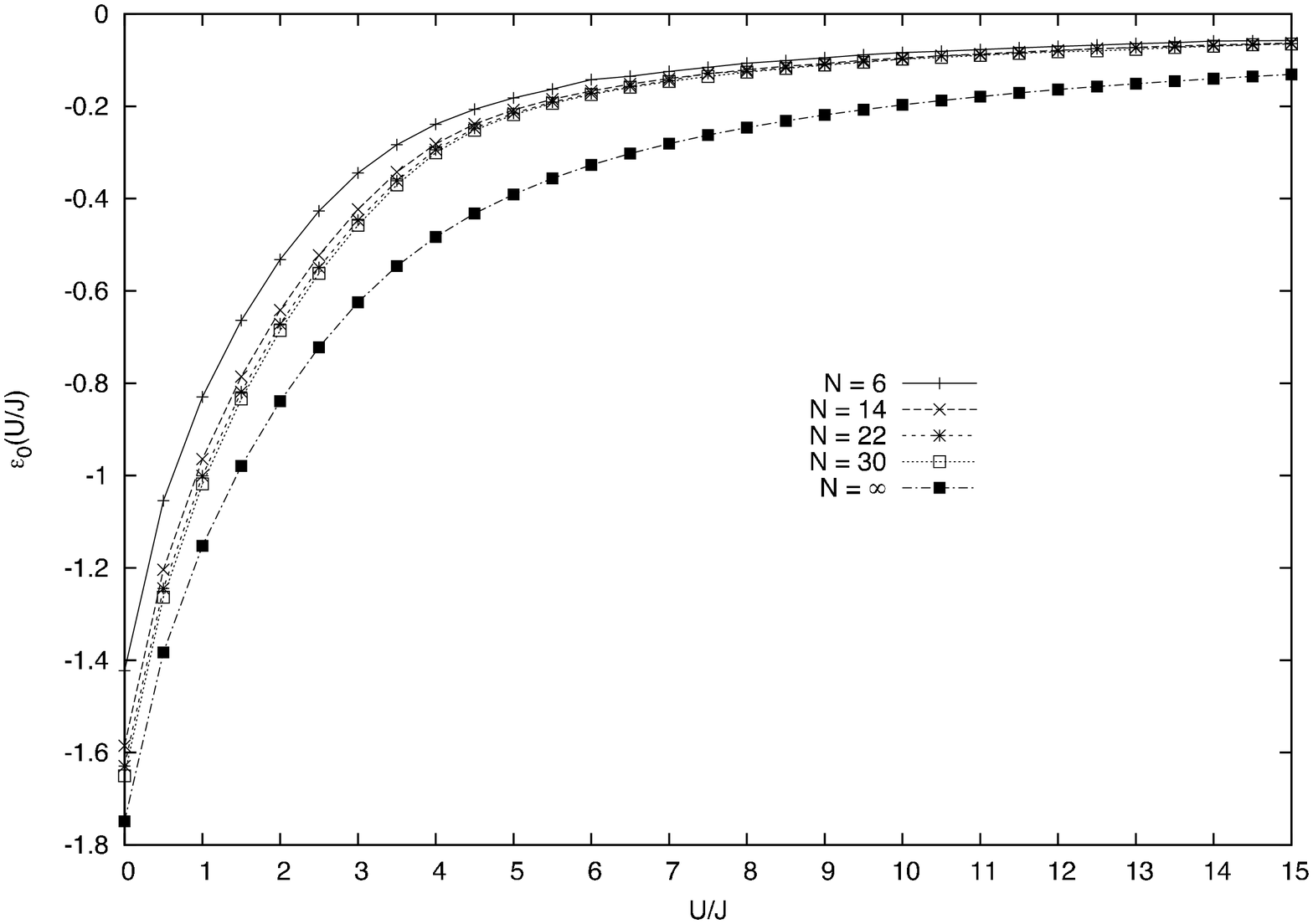}
\caption{Energy per lattice site of the ground state for the 1-dimensional quantum rotor model, for $d = 7$ and $\chi = 20$, as computed with MPS algorithms for finite $N$ and in the thermodynamic limit. Convergence with $N$ is seen to appear towards the limit $N \rightarrow \infty$.}
\label{energyqr}
\end{figure}

\begin{figure}
\includegraphics[width=.51\textwidth,angle=-90]{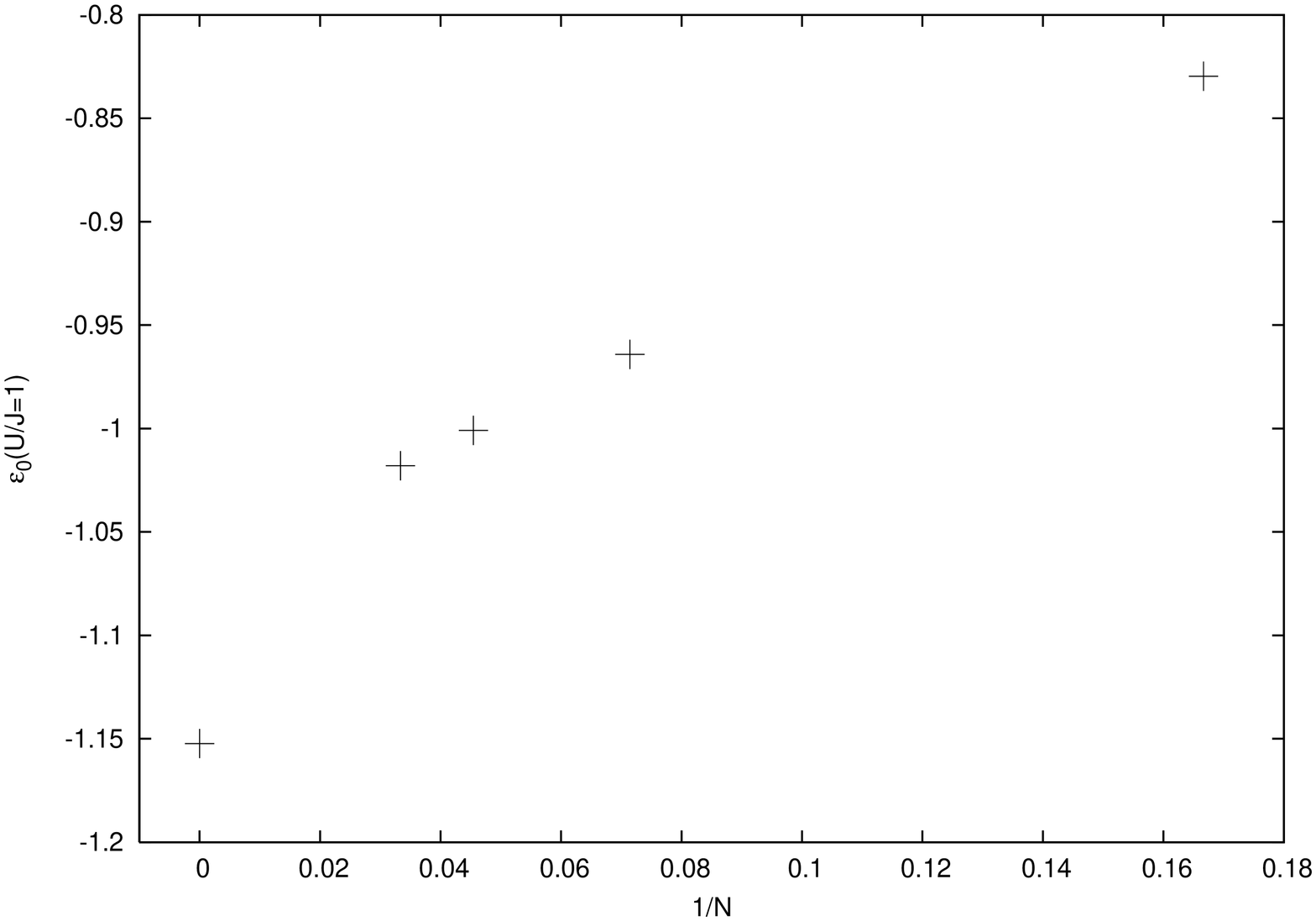}
\caption{Convergence with $N$ of the energy per lattice site of the ground state for the 1-dimensional quantum rotor model, for $d = 7$, $\chi = 20$ and $U/J = 1$, as computed with MPS algorithms for finite $N$ and in the thermodynamic limit.}
\label{convergence}
\end{figure}

We have performed computations of the ground state energy per lattice site of the system $\epsilon_0$, both in the finite-size regime and in the thermodynamic limit, as a function of the ratio $U/J$. Initially, we have performed our analysis for a local dimension $d=7$ and $\chi=20$. The results can be found in Fig.\ref{energyqr} and Fig.\ref{convergence}, which agree with the previous results from \cite{juanjo}, and where it is possible to see the convergence with $N$ of the ground state energy found with the algorithms for MPS with open boundary conditions and for MPS in the thermodynamic limit. 

\begin{figure}
\includegraphics[width=.7\textwidth]{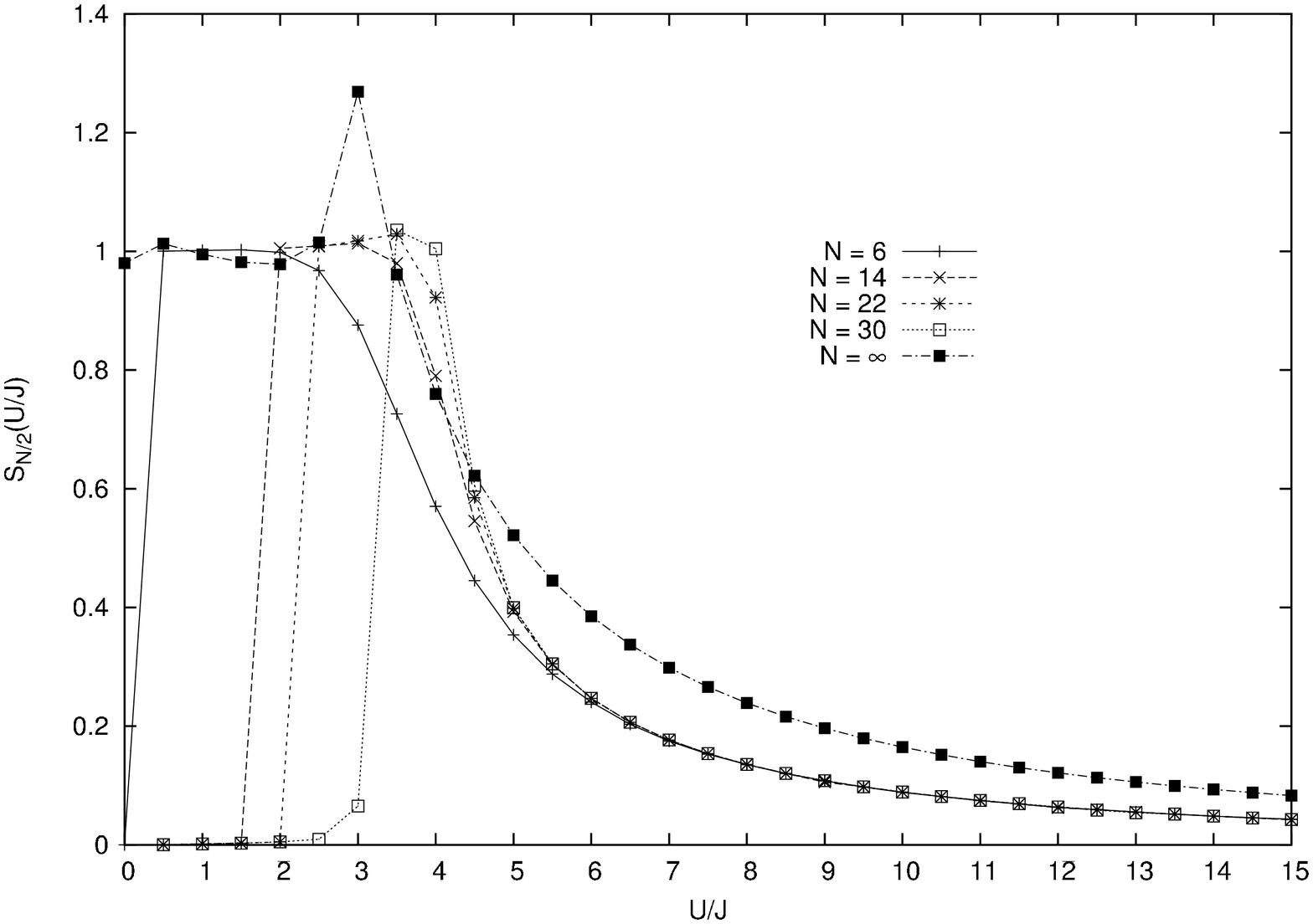}
\caption{Entanglement entropy of half a system for the 1-dimensional quantum rotor model, for $d=7$ and $\chi = 20$, as computed with MPS algorithms for finite $N$ and in the thermodynamic limit. Spontaneous symmetry breaking of the ${\mathbb Z}_2$ symmetry of the ground state is observed for the algorithms with finite $N$ towards the limit $U/J \rightarrow 0$, which is not observed in the infinite-size case. This makes the wave function to decay from a GHZ-like superposition of two states to a completely separable state, which results in a null value of the entanglement entropy.}
\label{entropyqr}
\end{figure}

In order to have a more accurate description of the properties of the ground state close to the critical point, we have performed as well a computation of the entanglement entropy of half a system, as a function of the ratio $U/J$, and again both in the finite-size case and the thermodynamic limit. Our results are represented in Fig.\ref{entropyqr}, which were obtained by using the same values in our MPS techniques than those used in the previous computation of the ground-state energy. According to the maximum observed value of the entanglement entropy, the quantum phase transition appears to be in our infinite-size simulations around $U/J \sim 3$, also in accordance with previous estimates from \cite{juanjo}. For finite $N$, our MPS numerical algorithms constantly exhibit spontaneous symmetry breaking of the ${\mathbb Z}_2$ symmetry of the ground state when approaching the limit $U/J \rightarrow 0$, which makes the entropy suddenly decrease towards $0$ (separable state). The MPS algorithm in the thermodynamic limit has shown to be much more stable in this respect, as it always maintains the GHZ-like superposition in the ground state when approaching the same limit, implying the entropy to evolve towards $1$. A divergence in the correlations is also seen to appear when increasing $N$ at the critical point. However, we can only hint such a divergence because of the finite value of $\chi$ that we use in our simulations. 

\subsection{Finite-size corrections to the critical ground state energy}

\begin{figure}
\includegraphics[width=.7\textwidth]{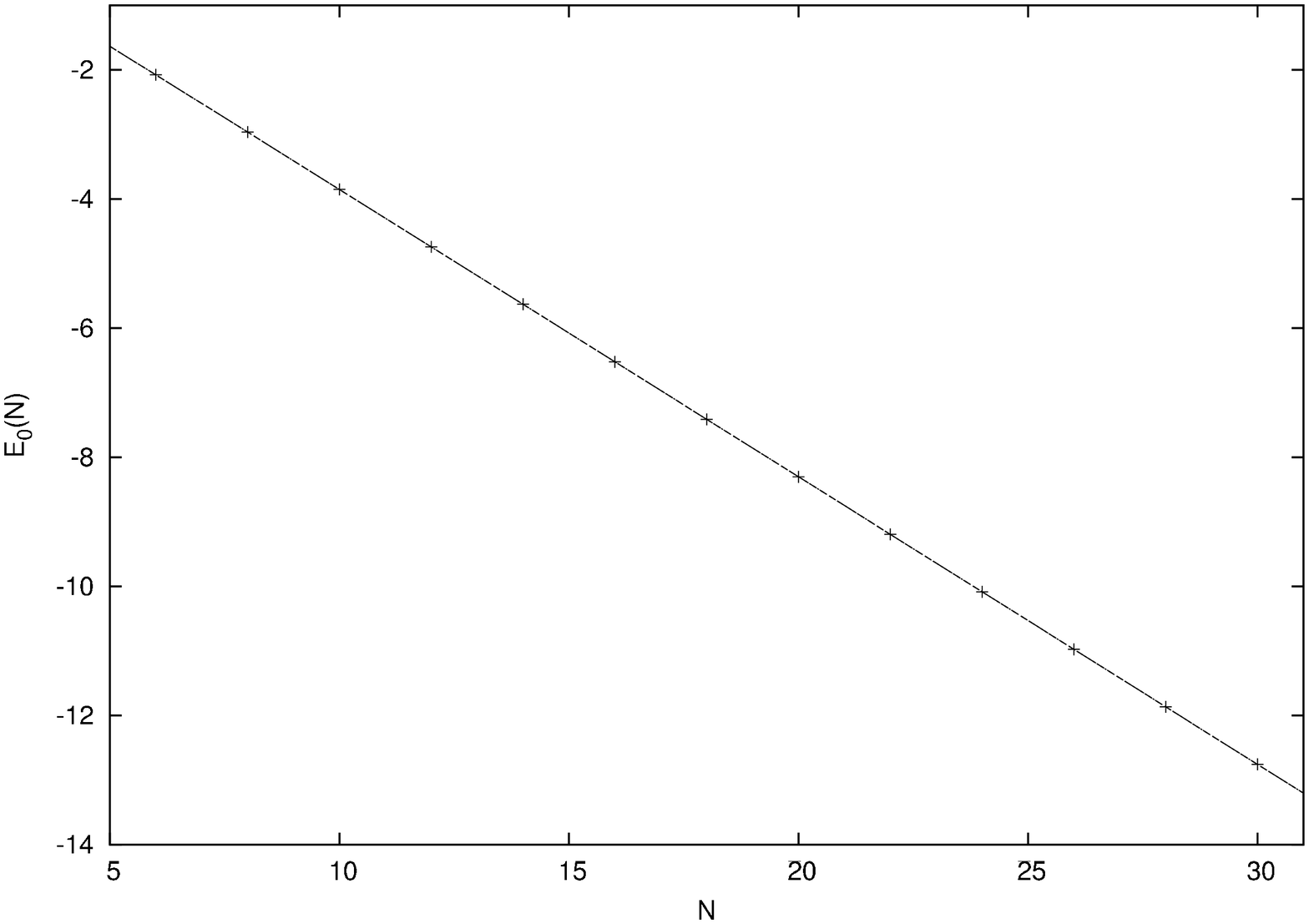}
\caption{Finite size scaling of the ground state energy at the point of maximum correlations for the 1-dimensional quantum rotor model, as computed with MPS algorithms with $d=14$ and $\chi = 16$. The line represents the fit to Eq.(\ref{cardyeq}), which provides a value of the central charge of $c=0.95 \pm 0.09$, which is compatible with the exact value $c=1$.}
\label{cardy}
\end{figure}

It is possible to make a check of consistency of the universality class of the model by studying the finite-size corrections of the ground state energy when close to the critical point. As proven in \cite{cardy}, for a system with finite size $N$ and open boundary conditions, the ground state energy approaches its critical value in the thermodynamic limit in the way dictated by the relation
\begin{equation}
E_0(N) = \epsilon_0(\infty) N + a - \frac{\pi c}{24} \frac{1}{N} + O\left(\frac{1}{N^2}\right)  , 
\label{cardyeq}
\end{equation}
where $\epsilon_0(\infty)$ is the energy per particle in the continuum limit, $a$ is a constant, and $c$ is the central charge describing the underlying universality class of the corresponding conformal field theory in the thermodynamic limit. Fitting the above expression to the ground-state energies computed at the value of maximum entropy for finite $N$ can then provide us with an estimation of the value for the corresponding central charge $c$. Our results for the 1-dimensional quantum rotor model can be found in Fig.\ref{cardy}, where we made use of finite-$N$ MPS algorithms with $d=14$ and $\chi = 16$. The fit to Eq.(\ref{cardyeq}) is good and provides a value of $c =  0.95 \pm 0.09$, which is to be compared with the exact known value $c=1$, corresponding to the $U(1)$-invariant critical point.    

\subsection{Spectrum of half an infinite system}

\begin{figure}
\includegraphics[width=.7\textwidth]{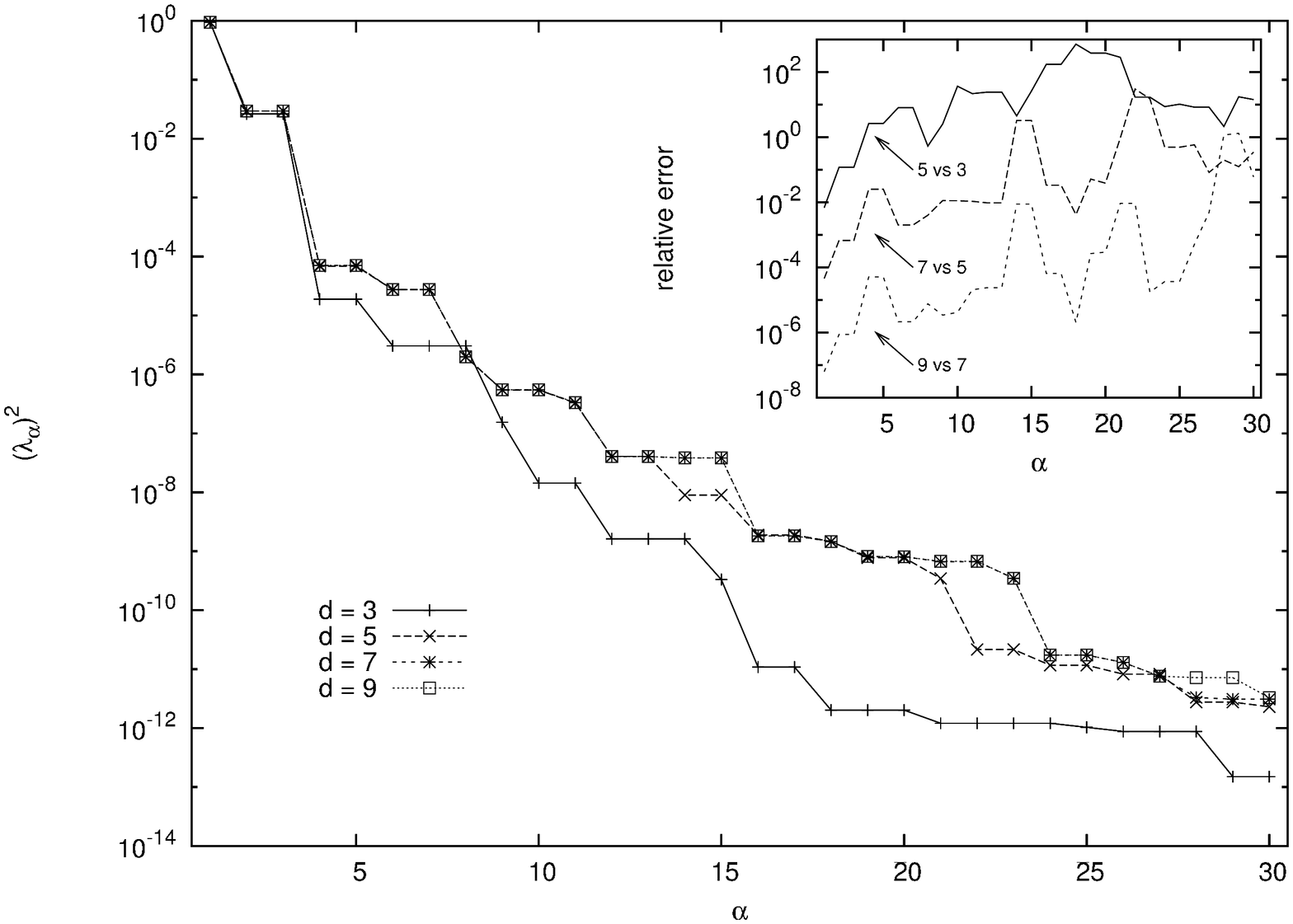}
\caption{Spectrum of the reduced density matrix of half an infinite system for the 1-dimensional quantum rotor model, as computed with MPS algorithms in the thermodynamic limit and $\chi = 30$. In the inset, convergence is seen to appear in the relative error of each one of the eigenvalues when increasing the local dimension $d$.}
\label{specconv}
\end{figure}

\begin{figure}
\includegraphics[width=.7\textwidth]{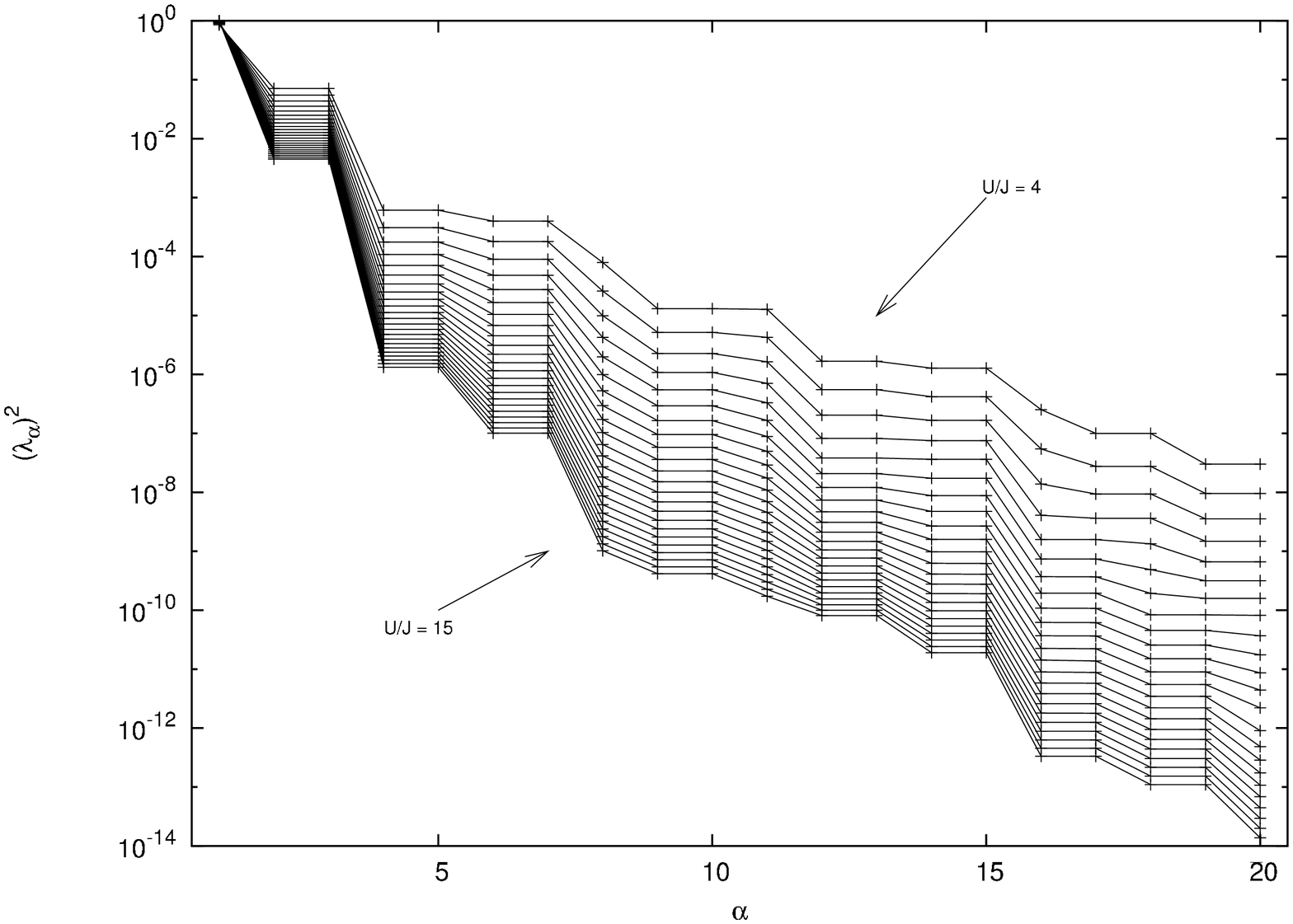}
\caption{Spectrum of the reduced density matrix of half an infinite system for the 1-dimensional quantum rotor model, as computed with MPS algorithms in the thermodynamic limit with $d=7$ and $\chi = 20$, when flowing away from criticality in the ratio $U/J$. A remarkable monotonicity in the spectrum is observed.}
\label{specuj}
\end{figure}

We have also performed a numerical evaluation in the thermodynamic limit of the spectrum of the reduced density matrix of half an infinite system. First, the convergence of the spectrum for finite $\chi = 30$ and off-critical $U/J$ has been considered as the local dimension $d$ increases. Our results are shown in Fig.\ref{specconv}, where it is possible to observe the convergence in the relative error of each one of the $\chi$ eigenvalues $(\lambda_{\alpha})^2$ as $d$ increases. This error is found to be of the order of $10^{-8}$. Second, we have performed a computation of the behaviour of the spectrum of the half-infinite reduced density matrix as a function of $U/J$ and away from criticality. In Fig.\ref{specuj} we show the found results for $d = 7$ and $\chi = 20$, where a remarkable monotonicity in the eigenvalues and their degeneracies when flowing away from criticality in the parameter space is appreciated. Indeed, it is possible to prove that such a monotonicity is very much related to the preservation of part of the conformal structure of the critical point, when flowing away from criticality in the space of parameters, involving in turn the monotonic majorization of the considered probability distribution \cite{rgr1,roman1, huan, romantoappear}. 

\section{Numerical solutions of partial differential equations}\label{sec:pdemps}

The fact that a set of coefficients can be approximated up to some accuracy by means of the contraction of a given tensor network (such as MPS), has applications well beyond the description of quantum many-body systems, as is the case of the MPS-assisted image compression \cite{photo}. In this section we show that the techniques previously introduced can also be used to deal with a purely mathematical problem, namely, that of finding numerical solutions to partial differential equations.  

Let $\Phi(x_1, \ldots, x_N)$ denote a function of $N$ real variables defined on some domain $\mathscr{D}$. Let $\hat{O}$ denote a linear differential operator, that is we assume that $\hat{O}$ is some polynomial in $x_1,\ldots, x_N,\partial_{x_1},\ldots, \partial_{x_N}$. The problem that we consider is to find a numerical solution, $\Psi(x_1, \ldots, x_N)$, of the equation
\beq\label{eq:pdeeq}
\hat{O} \Psi+\Phi=0, 
\eeq
on $\mathscr{D}$. We suggest that for wide classes of partial differential equations, using $N$-variable matrix product functions might turn a powerful computational method. The intuition underlying our point is most simple. A celebrated method to solve partial differential equations is to assume separation of variables, that is, that the solution has the form $\Pi_{i=1}^{N} \phi^{(i)}(x_i)$. In the language of quantum information, such a solution would correspond to a pure product state. On another hand, a fully general solution would have the form
\beq\label{eq:anyent}
\Psi(x_1, \ldots, x_N)=\sum_{s_1 \ldots s_N = 1}^{\infty}  c(s_1, \ldots, s_N) \phi^{(1)}_{k_1}(s_1) \ldots 
\phi^{(N)}_{s_N}(x_N).
\eeq
where the tensor $c(s_1, \ldots , s_N)$ is a priori not expected to have any particular structure. In the language of quantum information theory, Eq.(\ref{eq:anyent}) would correspond to an arbitrary pure entangled state, and could be, in principle, exponentially hard to describe as $N$ grows. Our point is that, in between, the finitely correlated MPS offer a much more refined description of the solution than the mere separation of variables, \emph{but} still allow for efficient numerical computations.

\subsection{The algorithm}

Let us work in what follows with periodic boundary conditions. In order to find a solution to Eq.(\ref{eq:pdeeq}), we will seek at minimising the error
\beq
W \equiv \int_{\mathscr{D}} \vert \hat{O} \Psi+\Phi \vert^2. 
\eeq

Now observe that if one sees the components of the matrices $A(k,s_k)$ related to the variable $k$ as those of a $d \chi^2$-component vector, $\gras{a}_k$, then $W$ appears as a positive quadratic form:
\beq\label{eq:errorpdequad}
W= \gras{a}_k^* \cdot R^{(k)} \cdot \gras{a}_k+\Gamma^{(k)^*} \cdot \gras{a}_k+ \gras{a}_k^* \cdot \Gamma^{(k)}+\int_\mathscr{D} \vert \Phi \vert^2,
\eeq
where $R^{(k)}$ is a $d \chi^2 \times d \chi^2$ positive matrix and $\Gamma^{(k)}$ is a $d \chi^2$ vector. Our algorithm, inspired from those used to find ground states of Hamiltonians described in the previous sections reads as follows: 

\begin{itemize}

\item Set the matrices $A(k,s_k)$ to some initial random values.

\item Sweep through each variable, $x_k$, sequentially, and improve the matrices $A(k,s_k)$ upon solving the linear system of equations 
\beq
R^{(k)} \cdot \gras{a}_k+\Gamma^{(k)}=0,
\eeq
until some desired convergence is attained.

\end{itemize}

At this point, it is worth informing the reader of three useful tricks that help to make the computations  well-behaved. First, the variables should always be "visited" in the \emph{same} order. For example, if the labels of the variables were arranged on a ring, the matrices related to each variable should always be improved in a clockwise order. Second, we have chosen the random matrices to which the MPS is initialised to be unitary (preconditioning of the solution). Third, after improvement of the matrices related to the $k$-th variable, we recommend to perform the gauge transformation described in \cite{vers04}:  we compute the singular decomposition $A(k,s_k)_{\alpha_{k-1},\alpha_k}=\sum_{\lambda,\mu=1}^{\chi} 
U^{(k)}_{(s_k \alpha_{k-1}), \lambda} \Sigma^{(k)}_{\lambda,\mu} W^{(k)}_{\mu,\alpha_k}$ where $U^{(k)\dagger}U^{(k)}=\mathbb{I}_{\chi}$, and make the gauge transformation $A(k,s_k)_{\alpha_{k-1},\alpha_k} \to U(k,s_k)_{\alpha_{k-1}, \lambda} \equiv U^{(k)}_{(s_k \alpha_{k-1}),\lambda}$.

\subsection{An example}

We have considered the Poisson equation:
\beq
\Delta \Psi+\Phi=0,
\eeq
where $\Delta=\sum_{i=1}^N \frac{\partial^2}{\partial x_i^2}$. The domain we have considered is $\mathscr{D}=[0,1]_1 \times \cdots \times [0,1]_N$. As individual set of functions, we have chosen the plane waves with integer wave index for all the possible sites:
\beq
\phi^{(k)}_{s_k}(x_k) \equiv \phi_{s_k}(x_k)=e^{i 2 \pi s_k x_k}, \hspace{0.7cm} s_k = -(d-1)/2, \ldots, (d-1)/2 \hspace{0.7cm} \forall k.
\eeq

We choose $\Phi$  to be of the MPS form.
\beq\label{eq:mpsPhi}
\Phi(x_1, \ldots, x_N)=\sum_{s_1 \ldots s_N=-(d-1)/2}^{(d-1)/2} \tr(F(1,s_1) \ldots F(N,s_N))\phi_{s_1}(x_1) \ldots \phi_{s_N}(x_N),
\eeq
where the matrices $F$ have size $\chi_{\Phi} \times \chi_{\Phi}$. Note that taking $\chi_{\Phi}$ and $d$ large enough, any (smooth) function $\Phi$ can be represented in the form (\ref{eq:mpsPhi}).

We have performed three different sets of computations for $N=8,20$ and $40$ variables. All our computations are such that $\chi_{\Phi}=d$, and such that the matrices $F$ are chosen to be random and unitary. To fairly evaluate the performance of our algorithm, we define a relative error $\delta$ as
\beq
\delta\equiv\frac{\int_{\mathscr{D}} \vert \hat{O} \Psi+\Phi \vert^2 }
{\int_{\mathscr{D}} \vert \hat{O} \Psi \vert^2+ \int_{\mathscr{D}} \vert \Phi \vert^2 
+2 \sqrt{\int_{\mathscr{D}} \vert \hat{O} \Psi \vert^2 \times \int_{\mathscr{D}} \vert \Phi \vert^2 } }
\eeq
This definition is inspired by the Cauchy-Schwarz inequality  ($0 \leq \delta \leq 1$). 

Our results are displayed in table \ref{tab:poisson8}. It is remarkable that the relative error decreases approximately exponentially with $\chi$.

\begin{table}
\begin{center}
\begin{tabular}{|c|c|c|c|c|c|c|}
\hline
 $N$ & $d$ &  $\chi_{\Phi}$ & $\chi$  & $W$ & $\delta$\\
\hline
\hline
$8$ & $3$ & $3$ & $3$&$3.25 \times 10{-3}$&$8.13 \times 10{-4}$\\
$8$ & $3$ & $3$ & $4$&$8.4 \times 10{-4}$&$2.08  \times 10{-4}$\\
$8$ & $3$ &$3$ & $5$&$1.36 \times 10{-4} $&$3.4 \times 10{-5}$\\
$8$ & $3$ & $3$ & $6$&$4.9 \times 10{-5} $&$ 1.23 \times 10{-5}$\\
$8$ & $3$ & $3$ & $7$&$3.4 \times 10{-5}$&$8.51 \times 10{-6}$\\
$8$ & $7$ & $7$ & $8$&$4.15 \times 10{-3}$&$1.04 \times 10{-3}$\\
$8$ & $7$ & $7$ & $9$&$3.44 \times 10{-3}$&$8.61 \times 10{-4}$\\
\hline
$20$ &$5$ & $5$ & $7$&$4.4 \times 10{-4}$&$1.1 \times 10{-4}$\\

$20$ &$7$ & $7$ & $3$&$4.09 \times 10{-4}$&$1.02 \times 10{-4}$\\
$20$ &$7$ & $7$ &  $4$&$3.02 \times 10{-4}$&$7.56 \times 10{-5}$\\
$20$ &$7$ &  $7$ & $5$&$1.2 \times 10{-4}$&$3 \times 10{-5}$\\
$20$ &$7$ & $7$ & $6$&$9.73 \times 10{-7}$&$2.43 \times 10{-7}$\\
\hline
$40$ &$3$ &  $3$ & $3$ &$7.3 \times 10{-5}$&$1.84 \times 10{-5}$\\
$40$ &$3$ &  $3$ & $4$ &$6.17 \times 10{-5}$&$1.54 \times 10{-5}$\\
$40$ &$3$ &  $3$ & $5$ &$3.96 \times 10{-5}$&$9.9 \times 10{-6}$\\
$40$ &$5$ &  $5$ & $5$ &$1.58 \times 10{-4}$&$3.94 \times 10{-5}$\\
\hline
\end{tabular}
\end{center}
\caption{Performances of the MPS algorithm for solving a Poisson equation. $N$ denotes the number of variables, $d$ denotes the number of local levels, $\chi_{\Phi}$ is the dimension of matrices $F$, $\chi$ is the dimension of the matrices of the MPS ansatz, $W$ is the absolute error, and $\delta$ the relative error.}\label{tab:poisson8}
\end{table}

\subsection{Including boundary conditions}
We have shown how to solve an equation on a given domain without extra constraints. Therefore, the algorithm converges to \emph{some} solution. Still, one often demands to be able to take into account some boundary conditions. Suppose that they are given by a set of $m$ relations of the form
\beq
\mathscr{C}_i [\Psi(x_1,\ldots,x_N)]+ f_i(x_1,\ldots,x_N)=0 
\eeq
where the relation $i$ is expected to hold on some domain $\mathscr{D}_i, \; i=1,\ldots,m$. When 
$\mathscr{C}_i$ is again of the form poly$(x_1,\ldots,x_N,\partial_{x_1},\ldots,\partial_{x_N})$, then the boundary condition error 
\beq
W_{\textrm{BC}} \equiv \sum_i \int_{\mathscr{D}_i} \vert \mathscr{C}_i [\Psi(x_1,\ldots,x_N)]+ f_i(x_1,\ldots,x_N) \vert^2
\eeq
is of the form (\ref{eq:errorpdequad}) for some matrices $R_{\textrm{BC}}^{(k)}$, and some vectors
$\Gamma_{\textrm{BC}}^{(k)}$. So after making the transformation $R^{(k)} \to 
R^{(k)}+R_{\textrm{BC}}^{(k)}$, $\Gamma^{(k)} \to \Gamma^{(k)}+\Gamma_{\textrm{BC}}^{(k)}$, we see that, with our method, solving a partial differential equation with boundary conditions can be done in exactly the same way as without.

\subsection{Discussion}

Let us now discuss the principal limitations of the methods. First, when the differential operator and the boundary conditions are not linear, the absolute error is no more of the convenient form 
(\ref{eq:errorpdequad}). More sophisticated methods are then necessary to perform the extremisation. 

The second limitation is related to the shape of the domain on which the variables are defined. As a premise to the algorithm, one should calculate tensors of the form
\beq
G^{j_1,\ldots,j_N}_{i_1,\ldots,i_N}(\mathscr{O}_1, \ldots, \mathscr{O}_N)=\int_{\mathscr{D}} 
\phi^*_{j_1}(x_1)  [\mathscr{O}_1 \phi_{i_1}](x_1)  \ldots \phi^*_{j_N}(x_N)
[\mathscr{O}_N \phi_{j_N}](x_N),
\eeq
where $\mathscr{O}_1, \ldots, \mathscr{O}_N$ are operators related to the variables $1,\ldots,N$ respectively. 

In the example we have treated, the domain was a mere Cartesian product of sub-domains, which had allowed to break $G$ into pieces $G^{j_1,\ldots,j_N}_{i_1,\ldots,i_N}(\mathscr{O}_1, \ldots, \mathscr{O}_N)= g^{j_1}_{i_1}(\mathscr{O}_1) \ldots g^{j_N}_{i_N}(\mathscr{O}_N)$. But for an arbitrarily shaped domain, the computational resources (time and memory) necessary to compute $G$ could grow exponentially with $N$. Still the method is not limited to the simplest case where the whole domain could be factored as a product of independent ones. For example, in the case of problems involving relatively small number of variables, this limitation is not so constraining. Second, when $\mathscr{D}$ can be decomposed as
\beq
\mathscr{D}= \bigcup_i \mathscr{D}_{i,1} \times \ldots \times \mathscr{D}_{i,n_i},
\eeq
where each $\mathscr{D}_{i,j}$ involves a small number of variables, then $G$ can be decomposed accordingly, and thus computed efficiently. 
These first two comments hold for the boundary conditions as well.

The other drawbacks of the methods are those generally related to MPS algorithms. First, we have no guarantee that the solution provided is a global minimum. However, contrarily to the study of Hamiltonian ground states, we have here a figure of merit that tells us how close to the actual solution the algorithm gets. The second drawback is related to the very structure of the ansatz. When studying many body quantum systems, it is known that the performance of MPS algorithms in their usual context depends on how far one is from criticality, performing well when the correlations between the particles are finite. In turn we expect our algorithms to behave similarly, and to perform  poorly for equations which solution are highly entangled functions. 

Despite its limitations, it is obvious that the method is very versatile and we believe that it can be successfully applied to a very wide class of differential problems. It certainly deserves further tests on relevant equations as well as theoretical investigation.

\section{Conclusions and outlook}\label{sec:conclusions}

In this paper we have presented in a pedagogical way several numerical optimization techniques based on MPS, both for finite systems (with open and periodic boundary conditions) and in the thermodynamic limit. We have shown how these algorithms can be extended in order to study quantum many-body systems with continuous degrees of freedom (continuous variables), by means of appropriate truncations in the local Hilbert spaces. On the one hand, and as a test, we have applied our algorithms to study a finite 1-dimensional chain of harmonic oscillators.  On the other hand we have performed a detailed numerical study of the properties of the 1-dimensional quantum rotor model. In both cases, we have found high precision results when working off criticality.  

The methods presented in our work can be further applied to the study of different continuous-variable quantum many-body systems. For example, it could be possible to consider the behaviour of anyonic excitations in small Bose-Einstein condensates \cite{pare01}, by means of a numerical treatment of Laughlin and Pfaffian wavefunctions \cite{laughlin1, laughlinnuestro, laughlinotro, ptaffian}. 
Also, it could be possible to perform a study of systems which alternate both discrete and continuous degrees of freedom, such as spin-boson models \cite{spinboson}. Interacting quantum field theories such as the $(1+1)$-dimensional $\lambda \varphi^4$ theory can also be studied by means of these techniques \cite{hc4}, and will be the main topic of a further work \cite{inpreparation}. 

Clearly, our work can also be extended to alternative tensor networks, such as the 2-dimensional PEPS \cite{vers04b} or the MERAs \cite{vida05}. The most delicate point when applying MPS algorithms is probably the choice of a basis. In the case we have treated, the choices (Fock states or plane waves) came quite naturally. But more often than not, to guess a correct single-particle basis in which truncate the Hamiltonian, is not an easy business. 

Finally, we have exhibited an MPS-like method to solve partial differential equations. The results we have got with the example treated are quite promising, and we hope that the method will find many applications. Further studies of it are currently under investigation.

\section{Acknowledgements}
We thank J.J. Garc\'ia-Ripoll for helpful discussions and for originally drawing our attention to the quantum rotor model. We also acknowledge MEC (Spain), QAP (EU),  Grup consolidat (Generalitat de Catalunya) and The University of Queensland (Australia) for financial support.

{}

\end{document}